\documentclass[a4paper,11pt]{article}
\def\be#1\ee{\begin{equation}#1\end{equation}}
\def\ba#1\ea{\begin{align}#1\end{align}}

\newcommand{\lr}[1]{\left( #1\right)}
\newcommand{\exv}[1]{\left< #1\right>}

\newcommand{\dd}{\mathrm{d}}
\renewcommand{\vec}[1]{\mathbf{#1}}

\newcommand{\eqqref}[1]{\text{eq.}\,\eqref{#1}}
\newcommand{\intbulk}{\int\dd^5 x\sqrt{-g}}
\newcommand{\intbdy}{\int\dd^4 x\sqrt{-h}}
\newcommand{\vielbein}[2]{e^{#1}_{\underline{#2}}}

\pdfoutput=1 % if your are submitting a pdflatex (i.e. if you have
\usepackage{jheppub} % for details on the use of the package, please
\usepackage{subcaption}

\usepackage[T1]{fontenc} % if needed
\usepackage{slashed}

\title{\boldmath Massive Dirac fermions from holography}

\author{N.W.M. Plantz,}
\author{F. Garc\'{i}a Fl\'{o}rez}
\author{and H.T.C. Stoof}
\affiliation{Institute for Theoretical Physics, Utrecht University, \\Princetonplein 5, 3584 CC Utrecht, The Netherlands}

\emailAdd{n.w.m.plantz@uu.nl}
\emailAdd{f.garciaflorez@uu.nl}
\emailAdd{h.t.c.stoof@uu.nl}

\abstract{We provide a framework to compute the dynamics of massive Dirac fermions using holography. To this end we consider two bulk Dirac fermions that are coupled via a Yukawa interaction and propagate on a gravitational background in which a mass deformation is introduced. Moreover, we discuss the incorporation of this approach in semiholography. The resulting undoped fermionic spectral functions indeed show that the Yukawa coupling induces a gap in the holographic spectrum, whereas the semiholographic extension is in general gapped but additionally contains a quantum critical point at which the effective fermion mass vanishes and a topological phase transition occurs. Furthermore, when introducing doping, the fermionic spectral functions show a quantum phase transition between a gapped material and a Fermi liquid.}

\begin{document} 
\maketitle
\flushbottom

\section{Introduction}
\label{sec:intro}
The holographic principle has established itself as a common instrument for the description of strongly coupled systems. While originally applied to supersymmetric theories such as $\mathcal{N}=4$ super-Yang-Mills theory \cite{Maldacena1999,Witten1998,Gubser1}, it was soon realized that the correspondence could be used to model real-world systems as well, such as quantum chromodynamics and the quark-gluon plasma \cite{Policastro1,Kovtun1}. In addition, over the past decade holography has been extended to include condensed-matter theory, which has led to the description of many strongly coupled condensed-matter phenomena by means of weakly coupled gravitational theories \cite{HHH1,HartnollLec,McGreevy2009,Sachdev2011}. These descriptions provide a great tool to compute thermodynamic and hydrodynamic properties, and besides this, also the spectra of bosonic or fermionic operators that are present in the dual condensed-matter field theory. 

Condensed-matter systems are usually described by nonrelativistic Dirac fermions. In holography, a common approach to cope with nonrelativistic systems is to use a Lifshitz background, which leads to a dynamical scaling exponent $z$ in the boundary theory that is different from its value in relativistic theories, i.e.,  $z\neq1$ \cite{Son1,McGreevy1,Kachru1,Taylor1}.  Alternatively, an asymptotically anti-de Sitter gravity theory may have an emergent infrared (IR) Lifshitz geometry with a scaling exponent $z$ different from $1$, so that the dynamics obtained from such theories can be reminiscent of nonrelativistic physics when restricted to the long-wavelength and low-frequency limit. However, Lifshitz backgrounds generally yield gapless particle-hole symmetric spectra and in both of these approaches, a missing ingredient is a Dirac mass in the spectrum. Such models are therefore great candidates for the description of effectively massless systems, such as single- or bilayer graphene, or the more recently discovered Dirac and Weyl semimetals \cite{Sachdev1,HartnollLec,JacobsUndoped}. For other purposes, it is desirable to extend the holographic model to also be able to describe spectral functions of massive fermionic operators that are ubiquitous in condensed matter. In this work we study the fermionic spectral functions that are obtained from such an extension.

To see what such an extension entails, it is important to realize that the reason that the fermionic spectra obtained from holography are in general gapless is twofold. Firstly, by introducing a probe Dirac spinor in the gravitational bulk theory, the fermionic spectral function on a boundary theory corresponds to a chiral fermion and is therefore massless \cite{Henningson1,Lee1,Vegh1,Vegh2,Zaanen1}. Secondly, introducing a mass in the boundary theory requires introducing a new scale in the conformal field theory (CFT), which implies that it is necessary to add a deformation to the bulk. The latter deformation was introduced in refs. \cite{Landsteiner1,Landsteiner2}, which focused on a model for the conductivity of a topological Weyl semimetal. In this paper, the model used to obtain the fermionic spectral functions includes such a deformation, as well as an additional Dirac spinor in the bulk, yielding the required amount of degrees of freedom on the boundary to describe a Dirac fermion. A model with two Dirac fermions in the bulk has already been used in ref. \cite{Jacobs1} to study Dirac semimetals. In this work we additionally introduce a coupling of the two Dirac fermions in the bulk to provide a coupling between the chiral fermions on the boundary, which is necessarily present for massive fermions. A similar construction was very recently described in ref. \cite{Liu1}, which appeared while completing this paper, where the approach was used to study semimetals with nodal lines. We would like to stress that our emphasis here is not on Weyl or nodal-line semimetals, but more generally on the description of fermionic spectra in condensed-matter systems which generally contain a Dirac mass. In these spectra, this mass can for instance be interpreted as an effective mass or gap in a band structure, which is the viewpoint taken here. However, an alternative viewpoint of the framework we present could be to interpret this mass as a real particle mass. This could then serve as a starting point for a holographic description for e.g. strongly coupled ultracold Fermi gases, which contain massive atoms.

In experiments we are usually interesed in single-particle spectral functions rather than the correlation functions of a composite fermion that are typically obtained in holography. Such single-particle spectral functions can be obtained from semiholography \cite{Faulkner1,Arpes}. Therefore, this paper also covers the incorporation of the aformentioned extension to massive Dirac fermions in a semiholographic framework. 

This paper is organized as follows. In section 2, we firstly present the procedure to obtain the Dirac fermion dynamics from holography. This means that we first specify a suitable gravitational background and then present the equations corresponding to the probe fermions propagating on top of this background. Moreover, in this section we also outline the procedure to obtain both the holographic and the semiholographic Green's functions. We present our results in section 3, where we compute the fermionic spectra using numerical solutions to the equations presented in section 2. Concluding in section 4, we discuss our results and comment on possible future directions.
\section{Obtaining massive Dirac fermions from holography} \label{sec:proc}
In this section we outline the procedure that we follow to obtain the dynamics of a Dirac fermion with a Dirac mass from holography. This procedure basically consists of solving two sets of coupled differential equations. We start by describing the first set, which gives us the gravitational bulk background that fixes quantities such as the temperature and chemical potential in the boundary field theory. We then derive the second set of differential equations, which describes the propagation of probe fermions in this bulk and gives us the holographic Green's function in the boundary theory. Finally, we derive an expression for the semiholographic Green's function of the Dirac fermion. To this end we use a dynamical-source model which is very similar to the one described in \cite{Arpes}, where the semiholographic Green's function for a chiral fermion is derived.

We refer to appendix \ref{app:conv} for conventions on the Dirac theory and the dimensionless units. Moreover, we always work in $d=4$ spatial dimensions in the bulk, implying that we consider a three-dimensional system on the boundary.
\subsection{Gravitational theory}
We wish to study a boundary theory containing massive Dirac fermions at nonzero chemical potential. As is well known, we can introduce the chemical potential by adding a $U(1)$ gauge field $A_\mu$ to the bulk \cite{HartnollLec}. As in ref. \cite{Landsteiner1}, we describe the mass deformation by adding a scalar field $\phi$ to the bulk. The mass of $\phi$ is fixed to $m_\phi^2=-3$, such that the operator dual to $\phi$ has dimension $\Delta=2+\sqrt{4+m_\phi^2}=3$. This agrees with the dimension of the operator $\exv{\bar{\psi}\psi}$ in a free boundary theory. Hence, the dimensions of the resulting deformation of the boundary theory match the dimensions of a free fermionic mass deformation $M_\psi\bar{\psi}{\psi}$. We discuss the choice of the mass $m^2_\phi$ in more detail in section \ref{sss:interp}. The source of the scalar field then acts as a Dirac mass $M_\psi$ on the boundary. The gravitational background we use therefore follows from the action:
\be \label{eq:ActionBR}
S_\text{background}=\intbulk\lr{R+12-\frac{1}{4}F^2- \frac{1}{2}\lr{(\partial\phi)^2+m_\phi^2\phi^2}}\,.
\ee
Considering static solutions with planar symmetry, we write the {\it Ansatz} for the metric as 
\be \label{eq:metricAnsatz}
\dd s^2=-f(r)e^{-\chi(r)}\dd t^2+\frac{\dd r^2}{f(r)}+r^2 \dd \vec{x}^2\, ,
\ee
where $(t,r,\vec{x})$ denotes the spacetime position in the bulk. Moreover, we use a temporal gauge field $A=A_t (r)\dd t$ and $\phi=\phi(r)$ due to planar symmetry. The coordinate $r$ is such that the black-brane horizon is at $r=r_+$, where $f(r_+)=0$, and the boundary is at $r=\infty$. The Hawking temperature is then given by 
\be \label{eq:HawkT}
T = \frac{f'(r_+)e^{-\chi(r_+)/2}}{4\pi}\,
\ee
and gives the temperature of the boundary theory. The equations of motion describing the background theory are
\begin{align}
\phi''+\lr{\frac{f'}{f}+\frac{3}{r}-\frac{\chi'}{2}}\phi'+\frac{3}{f}\phi &=0 \label{eq:KG}\, ,\\
A_t''+\lr{\frac{3}{r}+\frac{\chi'}{2}}A_t' &=0 \label{eq:MW}\, ,\\
\chi'+\frac{r}{3}\phi'^2 &=0 \label{eq:EE1}\, ,\\
f'+\lr{\frac{2}{r}-\frac{\chi'}{2}}f+\frac{r}{6}e^\chi A_t'^2-\frac{r}{2}\phi^2-4r &=0\, . \label{eq:EE2}
\end{align}
Notice that this background is very similar to those used to describe the holographic superconductor \cite{HHH2}, with a fixed bulk scalar mass and an uncharged bulk scalar. Therefore, following the arguments in ref. \cite{HHH2}, a solution to these equations is again determined by two initial conditions at the horizon $r_+$, namely $\phi(r_+)$ and $A_t'(r_+)$, assuming  $A_t(r_+)=0$. Moreover, using the following symmetry of the equations of motion, 
\be
\label{eq:scalerp}
r\rightarrow ar, \qquad (t,\vec{x}) \rightarrow (t,\vec{x})/a, \qquad \qquad f \rightarrow a^2 f, \qquad \qquad A_t \rightarrow a A_t,
\ee
we can put $r_+=1$. However, in contrast to the holographic superconductor, the solutions we consider here will also have a fixed nonzero scalar source term $\phi_s$, which is dual to the Dirac mass on the boundary. This means that both initial conditions that determine the background remain free, since we do not have to shoot for a solution without a source. The background is then described by two parameters, which are any two dimensionless ratios formed with the temperature $T$, the source $\phi_s$ and the chemical potential $\mu$ per unit charge, which follows from the boundary value of $A_t$.

One may wonder what happens with the instability that causes the phase transition for the holographic superconductor. A condition for this instability of the Reissner-Nordstr\"{o}m solution (with $\phi=0$) against the spontaneous formation of scalar hair is given by \cite{Horowitz0}
\be
q_\phi^2>\frac{m^2_\phi}{2}+\frac{d(d-1)}{8}.
\ee
Since in our case $d=4$, $m^2_\phi=-3$ and $q_\phi=0$, we do not satisfy this condition. Hence we do not expect this instability to occur, so that a solution with a nontrivial scalar profile should always have a nonzero source term. 
\subsection{Dirac Fermions}
We can calculate fermionic Green's functions by having probe Dirac fermions propagate on the fixed background described in the previous section. This is similar to the procedure presented in refs. \cite{Vegh1,Arpes}. However, the resulting fermionic Green's functions on the boundary correspond to a chiral fermion. The reason is that the Dirac equation in the bulk imposes a relation between the two chiral components of the probe fermion on the boundary. Let us quickly review this case. Denoting the probe fermion by $\psi$, we define the components 
\be
\psi_{R,L}=\frac{1}{2}\lr{1\pm \Gamma^{\underline{r}}}\psi, \qquad \psi_L = \begin{pmatrix} 0 \\ \psi_- \end{pmatrix}, \qquad \psi_R = \begin{pmatrix}  \psi_+ \\ 0 \end{pmatrix},
\ee
where the plus (minus) sign corresponds to $\psi_R$ ($\psi_L$). Note that $\psi=\psi_R+\psi_L$, whereas $\psi_{\pm}$ are two-component spinors with definite chirality on the boundary. We then add the following action to the bulk:
\be \label{eq:actionchiral}
S_{\text{Weyl}}=ig_f \intbulk \bar{\psi}\lr{\slashed{D}-M}\psi+ig_f\intbdy \bar{\psi}_R\psi_L.
\ee
Here $M$ is the bulk Dirac mass, $g_f$ is a coupling constant, $D_\mu=\nabla_\mu-iqA_\mu$ with $\nabla_\mu$ the spinor covariant derivative and $q$ the fermion bulk charge, so that the chemical potential of the spinor is $\mu=q A_t(\infty)$. The boundary action is included to make the variational principle well defined, and is consistent with the Dirichlet boundary condition $\delta\psi_R=0$. As shown in ref. \cite{Arpes}, writing out the Dirac equation $\lr{\slashed{D}-M}\psi=0$ in chiral components reveals that the relation between them can be written in the form
\be \label{eq:chiralxi}
\psi_-(r,k)=-i\xi(r,k)\psi_+(r,k),
\ee
where we Fourier transformed the spinors on slices of constant $r$. As a consequence, the action \eqref{eq:actionchiral} evaluated on shell can be written as
\be \label{eq:Soschiral}
S_{\text{Weyl}}^{\text{on shell}}=-ig_f\int_{r=r_0} \frac{\dd^4k}{(2\pi)^4}\sqrt{-h}\psi^\dagger_+\psi_-=-g_f\int_{r=r_0} \frac{\dd^4k}{(2\pi)^4}\sqrt{-h}\psi^\dagger_+\xi \psi_+.
\ee
Here $r_0$ is a cut-off surface, which as we shall explain later is important when computing a Green's function. Ultimately, we take the limit of $r_0$ going to infinity. From the above action it is clear that $\xi$ is proportional to the holographic Green's function for the chiral boundary operator that is sourced by the boundary value of the chiral spinor $\psi_+$. In other words, the chiral component $\psi_R$ of $\psi$ acts as a source for the chiral operator whose expectation value is contained in $\psi_L$, so that after integrating out $\psi_L$ we are left with the effective action for a chiral fermion. As described in e.g. ref. \cite{Arpes}, from the Dirac equation we can then derive a differential equation for $\xi$. Solving this using infalling boundary conditions, the holographic retarded Green's function for the chiral operator $O$ that couples to $\psi_+$ then follows from
\be
G_O(k)=\lim_{r_0\rightarrow\infty} r_0^{2M} \xi(r_0,k).
\ee
In the procedure above we have seen that we have to integrate out half of the degrees of freedom of the probe fermion. Therefore, in order to describe a Dirac fermion on the boundary, we double the amount of degrees of freedom by introducing two bulk fermions $\psi^{(1)}$ and $\psi^{(2)}$. Our goal is then to derive an effective action similar to equation \eqref{eq:Soschiral}, but this time with four-component spinors. Using the Dirichlet boundary conditions $\delta\psi^{(1)}_R=0$ and $\delta\psi^{(2)}_L=0$, we can derive such an effective action that contains the two chiral fermions $\psi^{(1)}_+$ and $\psi^{(2)}_-$. In order to describe a massive Dirac spinor, we also need to couple these chiral components. We do this by introducing a Yukawa interaction in the bulk, that couples the two fermions to the scalar field with coupling constant $g_Y$. The total action, including the boundary terms consistent with the abovementioned Dirichlet boundary conditions, then looks as follows:
\ba
S_{\text{Dirac}}=&\,\,ig_f\intbulk \lr{\bar{\psi}^{(1)}\lr{\slashed{D}-M}\psi^{(1)}+\bar{\psi}^{(2)}\lr{\slashed{D}+M}\psi^{(2)}} \nonumber \\
&+ig_Y\intbulk \phi\lr{\bar{\psi}^{(1)}\psi^{(2)}+\bar{\psi}^{(2)}\psi^{(1)}} +ig_f\intbdy \lr{ \bar{\psi}^{(1)}_R\psi^{(1)}_L-\bar{\psi}^{(2)}_L\psi^{(2)}_R}. \label{eq:action2}
\ea
Note that we took the mass of $\psi^{(2)}$ to be $-M$ so that the asymptotic behaviors of the sources  $\psi^{(1)}_R$ and $\psi^{(2)}_L$ are equal \cite{Jacobs1}. The equations of motion following from this action are
\ba 
\lr{{\slashed{D}-M}}\psi^{(1)}&=-\lambda\phi\psi^{(2)}, \label{eq:Dirac1}\\
\lr{{\slashed{D}+M}}\psi^{(2)}&=-\lambda\phi\psi^{(1)},\label{eq:Dirac2}
\ea
where $\lambda=g_Y/g_f$. Notice that without the Yukawa term, we would just end up with two copies of eq. \eqref{eq:Soschiral} and therefore describe two uncoupled chiral fermions. The corresponding Green's function would then be ungapped, and could therefore not correspond to the Green's function of the fermions that appear in the Dirac mass deformation that we introduced by adding the scalar field to the background. Hence a term such as the Yukawa term is necessary if we want to describe the dynamics of the Dirac fermion at the boundary. There may however be other possibilities to couple the two chiral components. This Yukawa term has the additional advantage that it does not change scaling dimensions of the operators dual to the bulk spinors. 

Let us now define the two bulk Dirac spinors $\Psi\equiv \psi^{(1)}_R+\psi^{(2)}_L$ and $\eta\equiv \psi^{(1)}_L-\psi^{(2)}_R$, i.e.,
\be
\Psi = \begin{pmatrix} \psi^{(1)}_+ \\ \psi^{(2)}_- \end{pmatrix}, \qquad \qquad \qquad \eta = \begin{pmatrix}  -\psi^{(2)}_+ \\ \psi^{(1)}_- \end{pmatrix}.
\ee
With our choice of the Dirichlet boundary conditions, $\Psi$ contains the sources. Similarly to the chiral case, we would then like to integrate out the other components that are contained in $\eta$, and derive an effective action for $\Psi$. Evaluating the action \eqref{eq:action2} on shell, the bulk terms vanish and we can write the boundary term as
\be
S^{\text{on shell}}_{\text{Dirac}}=ig_f\intbdy \bar{\Psi}\eta.
\ee
Rescaling the spinors to get rid of the spin connection, see the discussion around \eqqref{eq:rescale} in appendix \ref{app:Dirac} for details, the Dirac equation in momentum space can be written as
\ba
-\lr{e^r_{\underline{r}}\partial_r+M}\eta+\lr{i\slashed{\tilde{k}}+\lambda\phi}\Psi&=0, \label{eq:Diraceta}\\
\lr{e^r_{\underline{r}}\partial_r-M}\Psi+\lr{i\slashed{\tilde{k}}-\lambda\phi}\eta&=0. \label{eq:Diracpsi}
\ea
Here $\tilde{k}_\mu=(-(\omega+qA_t),\vec{k})$ so that the slash operator has no $r$-component. As in the chiral case, this imposes a relation between $\Psi$ and $\eta$ which can be written as
\be \label{eq:defXi}
\eta(r,k) = -i \Xi(r,k) \Psi(r,k).
\ee
The on-shell action then becomes
\be \label{eq:Seffholo}
S_{\text{Dirac}}^{\text{on shell}}=g_f\int_{r=r_0}\frac{\dd^4k}{(2\pi)^4}\sqrt{-h} \bar{\Psi}\Xi\Psi.
\ee
Here we can see that the $4\times 4$ matrix $\Xi$ is related to the Green's function for the fermionic operator $O$ that is sourced by the boundary value of the Dirac spinor $\Psi$. More precisely, using infalling boundary conditions that we specify later, the holographic Green's function is given by
\be \label{eq:holGR}
G_O(k)=-\lim_{r_0\rightarrow\infty} r_0^{2M} \Gamma^{\underline{0}}\Xi(r_0,k).
\ee
We will proceed by deriving a differential equation with which we can compute $\Xi$ directly. 
\subsubsection{Computing the holographic Green's function}
Using the Dirac equations in \eqref{eq:Diraceta} and \eqref{eq:Diracpsi}, we can derive a differential equation which we can solve for $\Xi$, thereby obtaining the holographic Green's function through \eqqref{eq:holGR}. Taking the derivative of \eqref{eq:defXi} gives
\be
\Xi e_{\underline{r}}^r\partial_r\Psi=ie_{\underline{r}}^r\partial_r \eta -e_{\underline{r}}^r\partial_r\Xi \Psi.
\ee
Multiplying \eqref{eq:Diracpsi} by $\Xi$ from the left and substituting the above then gives
\be
ie_{\underline{r}}^r\partial_r\eta-e_{\underline{r}}^r\partial_r\Xi \Psi-M\Xi\Psi+\Xi\lr{i\slashed{\tilde{k}}-\lambda\phi}\eta=0.
\ee
Eliminating $\partial_r\eta$ using  \eqref{eq:Diraceta} and $\eta$ using \eqref{eq:defXi} ultimately gives
\be \label{eq:Xieq}
\lr{-(e_{\underline{r}}^r\partial_r+2M)\Xi+i\lr{i\slashed{\tilde{k}}+\lambda\phi}-i\Xi\lr{i\slashed{\tilde{k}}-\lambda\phi}\Xi}\Psi=0.
\ee
This shows that we can compute $\Xi$ by solving the first-order nonlinear differential $4\times 4$ matrix equation between the brackets. However, we can greatly reduce the amount of equations we need to solve by exploiting rotational symmetry to set $k_\mu=\lr{-\omega,0,0,k_3}$. Using symmetry we can then write\footnote{This can also be shown by solving \eqref{eq:Diracpsi} for $\eta$ and reading off the matrix structure of $\Xi$.}
\be \label{eq:Xiexp}
\Xi=\Xi_0\Gamma^{\underline{0}}+\Xi_3\Gamma^{\underline{3}}+\Xi_c \mathbb{I}_{4}
\ee
where $\mathbb{I}_4$ is the $4\times 4$ identity matrix. The \textit{Ansatz} above in \eqref{eq:Xiexp} shows that there are only three degrees of freedom for which we have to solve. However, it is more insightful to write the equations in \eqqref{eq:Xieq} in terms of $\Xi_{\pm}\equiv \Xi_0 \pm \Xi_3$. This yields
\ba
(e_{\underline{r}}^r\partial_r+2M)\Xi_{\pm}&=\lr{\tilde{\omega} e^0_{\underline{0}}\mp k_3 e^3_{\underline{3}}}\lr{1-\Xi_c^2} + \lr{\tilde{\omega} e^0_{\underline{0}}\pm k_3 e^3_{\underline{3}}}\Xi_\pm^2+2i\lambda\phi\Xi_c \Xi_\pm, \nonumber \\
(e_{\underline{r}}^r\partial_r+2M)\Xi_c\,&=\lr{\tilde{\omega} e^0_{\underline{0}}+ k_3 e^3_{\underline{3}}}\Xi_+\Xi_c+\lr{\tilde{\omega} e^0_{\underline{0}}- k_3 e^3_{\underline{3}}}\Xi_-\Xi_c+i\lambda\phi\lr{1+\Xi_c^2-\Xi_+\Xi_-}, \label{eq:redXieqs}
\ea
where $\tilde{\omega}=\omega+q A_t$. As a check, notice that for $\lambda=\psi^{(2)}=0$ the lower-left $2\times 2$ block of $\Xi$ corresponds to the matrix defined in \eqref{eq:chiralxi} for the chiral case. From \eqref{eq:Xiexp} we see that the eigenvalues of this block are exactly $\Xi_\pm$. Setting $\lambda=\Xi_c=0$ in \eqref{eq:redXieqs} indeed recovers the equation for the chiral case, see e.g. eq. (2.31) in ref. \cite{Arpes}.

These equations can now be solved numerically to obtain the matrix $\Xi$. As they are first-order ODE's, we need to impose one initial condition for each component. Since only $e^0_{\underline{0}}$ diverges at $r_+$, we demand that in both equations the coefficient of this factor vanishes at the horizon. The second equation then yields either $\Xi_c(r_+)=0$ or $\Xi_+(r_+)=-\Xi_-(r_+)$. However, the latter is not consistent with the infalling boundary conditions, for which we know from the chiral case that the result is $\Xi_\pm(r_+)=i$. We conclude that we must impose $\Xi_c(r_+)=0$. The first equation then gives that $\Xi_\pm(r_+)=\pm i$, where the infalling boundary conditions require that we choose $+i$ for both cases.
\subsubsection{Obtaining the semiholographic Green's function}\label{sss:semhol}
Next, we use semiholography to derive an expression for the single-particle Green's function. We note that our approach is slightly different from the work in ref. \cite{Faulkner1}, where the authors use semiholography to capture universal IR physics. In contrast, our objective is to use semiholography to obtain the Green's function of an elementary fermion that is for instance measurable in ARPES experiments. To this end, we follow the approach outlined in ref. \cite{Arpes} for the chiral case, which is constructed such that the obtained semiholographic Green's function $G_R$ satisfies the sum rule that in our case reads
\be \label{eq:sumrule0}
\frac{1}{4\pi}\int_{-\infty}^\infty \dd \omega \,\text{Im Tr}\,G_R(\omega,\vec{k})= 1.
\ee
This procedure implies that we interpret the ultraviolet (UV) cut-off surface at a fixed radial coordinate $r=r_0$ as the boundary on which the single fermions live and interact with the CFT. In practice this means that the sources become dynamical and that the holographic Green's function derived above becomes the self-energy of the elementary fermion. In particular we note that the semiholographic Green's functions obtained in this manner are not restricted to IR physics, as the sum rule above also implies.

Above we have calculated the holographic contribution to the effective action, which is given by eq. \eqref{eq:Seffholo}. To this we add the free action on the UV brane for the source $\Psi$:
\be \label{eq:SUV}
S_{UV}=iZ\int_{r=r_0}\dd^4x \sqrt{-h}\bar{\Psi}\lr{\slashed{D}-\tilde{M}_0}\Psi.
\ee 
Note that we can add this action since we chose the Dirichlet boundary condition corresponding to $\delta\Psi=0$. The total effective boundary action can then be written as
\be
S_{\text{eff}}=\frac{Z\sqrt{-h}}{r_0}\int_{r=r_0}\frac{\dd^4 k}{(2\pi)^4}\bar{\Psi}\lr{-\Gamma^{\underline{a}}\tilde{k}_a-i\tilde{M}_0 r_0+\frac{g_fr_0}{Z}\Xi}\Psi,
\ee
where we Fourier transformed the fields and used that $e^0_{\underline{0}}\approx 1/r_0$ for $r_0$ near the boundary. Notice that the kinetic term now becomes canonically normalized upon rescaling the fields as $\Psi\rightarrow \sqrt{r_0/Z\sqrt{-h}}\Psi$. We can then take the following limit:
\be
r_0\rightarrow \infty, \quad g_f\rightarrow 0, \quad \tilde{M}_0 \rightarrow 0, \quad g\equiv \frac{g_fr_0^{1-2M}}{Z} = \text{const.}, \quad M_0\equiv  \tilde{M}_0 r_0 = \text{const.}
\ee
The effective action for the elementary Dirac fermion $\Psi$ can then be written as
\be
S_{\text{eff}}=\int\frac{\dd^4 k}{(2\pi)^4}\Psi^\dagger G_R^{-1}\Psi.
\ee
Here the inverse Green's function is given by
\be \label{eq:GRinvStandard}
G_R^{-1}(k)=\begin{pmatrix} \sigma\cdot \tilde{k} & iM_0\\ -iM_0 & -\bar{\sigma}\cdot \tilde{k} \end{pmatrix}+\Sigma
\ee
where $\sigma=(\mathbb{I}_2,\sigma^i)$ and $\bar{\sigma}=(-\mathbb{I}_2,\sigma^i)$ with $\sigma^i$ the Pauli matrices, and where we defined the self-energy
\be \label{eq:Sigmadef}
\Sigma(k)\equiv g \Gamma^{\underline{0}}\lim_{r_0\rightarrow\infty}r_0^{2M}\Xi(r_0,k).
\ee
Using again the  rotational symmetry to choose the momentum as $\tilde{k}_\mu=\lr{-\tilde{\omega},0,0,k_3}$ and using the notation of \eqref{eq:Xiexp}, we can also write the Green's function as
\be \label{eq:renGR}
G_R^{-1}(\omega,k_3)=\Gamma^{\underline{0}}\lr{\lr{\tilde{\omega}+g\Xi_0}\Gamma^{\underline{0}}+\lr{-k_3+g\Xi_3}\Gamma^{\underline{3}}+\lr{-iM_0+g\Xi_c}\mathbb{I}_4}.
\ee
This expression is evaluated at the boundary, so that $\tilde{\omega}=\omega+\mu$. We absorbed a factor $r_0^{2M}$ in the components of $\Xi$, such that these are finite at the boundary $r_0\rightarrow\infty$. The above expression shows that $\Xi_0$ is related to a wavefunction renormalization, whereas $\Xi_c$ acts like a mass renormalization.

When $g$ becomes large, the inverse Green's function will be dominated by the self-energy in \eqqref{eq:Sigmadef}. In this limit we recover the holographic Green's function, albeit rescaled by $1/g$ and in alternative quantization. The latter implies that this Green's function corresponds to the Dirichlet boundary conditions $\delta\psi_R^{(2)}=\delta\psi_L^{(1)}=0$, which gives the inverse of the Green's function in standard quantization.
\subsubsection{Interpreting semiholography} \label{sss:interp}
Before continuing to our results, we briefly comment on the physical picture we have in mind when applying the semiholographic procedure described above. On the one hand, we introduce free single fermions $\Psi$ living on a UV cut-off surface, located at $r_0$. On the other hand, we have a (deformed) CFT, containing a composite fermionic operator $O$.\footnote{In the description above, $O$ is the operator sourced by $\Psi$.} In essence, what happens in semiholography is that we linearly couple these two theories and subsequently integrate out the CFT part in order to obtain the effective Green's function of the fermion. In other words, we describe single fermions $\Psi$ interacting with a fermionic operator $O$ of the CFT. 

Since we are doing bottom-up holography, it is not known what the exact physical interpretation of $O$ is. A possible interpretation is to describe single fermions interacting with a completely unrelated composite fermionic operator. However, the physical picture we have in mind is a fermionic condensed-matter system, which at long wavelengths is described by a CFT with a number of collective variables of these fermions, e.g. electrons or atoms. In this case, the composite operator $O$ in the CFT `contains' the single fermion of interest, such as the electron or the atom.  Such a `self-consistent' interpretation imposes additional restrictions on our model. One example is the choice of the parameter $m_\phi^2$ in the bulk, which we then should indeed choose such that it describes a free-fermion mass deformation. This is because we know that the elementary fermion is described by a free theory in the far UV. Hence, if the fermion is a building block of the CFT, such a mass deformation should exist in the CFT. In contrast, if the CFT is unrelated to the fermion $\Psi$, we might as well have chosen a different value for $m_\phi^2$, as it is not clear that a mass-like deformation introduced by the dual field $\phi$ in the CFT should necessarily correspond to that of a free fermion.\footnote{See ref. \cite{Landsteiner3} for a holographic model where the value of $m^2_\phi$ is varied.}

In our model, both theories contain a mass scale. In the theory describing the elementary fermion, this mass is denoted by $M_0$. In the CFT the mass scale is set by the source $\phi_s$ of the scalar field $\phi$. The self-consistency requirement mentioned above means that also these two mass scales should be related, and enables us to fix the ratio $M_0/\phi_s$. An argument for this is given in appendix \ref{app:RG}. Although we use the fixed value presented there, i.e., $M_0/\phi_s=\sqrt[4]{\pi^2/3}$, we do not expect substantial qualitative differences in our semiholographic spectra when choosing a slightly different value for this ratio or for $m_\phi^2$ for that matter. 
\section{Fermionic spectral functions}
Given a Green's function $G(\omega,\vec{k})$, we can compute the spectral function defined as 
\be \label{eq:spf}
\rho(\omega,\vec{k})=\frac{1}{\pi}\text{Im Tr}\,G(\omega,\vec{k}).
\ee
Here we can take the holographic Green's function $G=G_O$ defined in \eqqref{eq:holGR} to obtain the spectral function of the fermionic operator $O$, or we can take $G=G_O^{-1}$ to obtain the holographic spectral function in alternative quantization. Alternatively, we can use the semiholographic Green's function $G_R$ defined in \eqqref{eq:GRinvStandard} to obtain the spectral function for the elementary fermion $\Psi$. We can think of this as moving away from the limit $g\rightarrow\infty$ which corresponds to the holographic Green's function in alternative quantization. It is however important to keep in mind that the holographic results can be obtained independently of semiholography. An important property of the semiholographic Green's function is that in contrast to the holographic one, it obeys the sum rule in \eqqref{eq:sumrule0}, i.e.,
\be \label{eq:sumrule}
\int_{-\infty}^\infty \dd \omega \rho(\omega,\vec{k})= 4.
\ee
This implies that $G_R$ is indeed the Green's function of an elementary Dirac fermion, which is a measurable quantity that contains the information about the spectrum of the fermion dynamics. Moreover, this property allows us to compute momentum distribution functions. The spectral function is normalized such that the sum rule gives the number of degrees of freedom, i.e., 4 for a Dirac fermion.

Before presenting our results, we take a moment to stress on which parameters the spectral functions depend. For the holographic spectral function this means we should state on which parameters the function $\Xi(\omega,\vec{k})$ depends. Firstly, this matrix depends on the gravitational background, which is specified by the scales $T$, $\mu$,\footnote{Actually, $\mu=q A_t(r=\infty)$ which also depends on $q$, but this dependence is trivial.} and $\phi_s$. Besides this, the self-energy depends on the parameters $\lambda$, $q$ and $M$, which all appear in \eqqref{eq:Xieq}. Here, $\lambda$ describes the coupling strength between the chiral components of the fermion. Therefore, it is necessarily nonzero for fermions with a mass term. Moreover, the bulk charge $q$ and bulk mass $M$ of the probe fermions are dimensionless parameters that define the CFT in which we calculate the two-point function.

For the semiholographic spectral function, $M$ is restricted to the range $|M|<1/2$. This is necessary for the sum rule and the Kramers-Kronig relations to hold \cite{Arpes}. The semiholographic spectral function depends on all the parameters above and in addition on $g$ and $M_0$ through \eqqref{eq:renGR}. The parameter $g$ is nonnegative and describes the strength of the coupling between the fermions $\Psi$ and the CFT, i.e., the strength of the self-energy. Here the limit $g\rightarrow\infty$ recovers the holographic Green's function in alternative quantization, whereas $g=0$ corresponds to a free massive Dirac fermion. The mass scale $M_0$ is fixed by $\phi_s$ as explained in section \ref{sss:interp} and in appendix \ref{app:RG}. In this work we scale all dimensionful quantities with $M_0$. This means that from this point on all quantities we refer to are implicitly scaled by the appropriate power of $M_0$ to make them dimensionless.\footnote{E.g. instead of $T/M_0$ we say $T$.} We fix $q=1$ and $M=1/4$ unless stated otherwise. We expect that changing these values should result mostly in quantitive rather than qualitative differences in the spectra. In this paper we mainly focus on the low-temperature case $T=1/100$, unless stated otherwise. 
\subsection{Undoped spectra} \label{ss:undoped}
First of all, we concentrate on the undoped case, i.e., $\mu=0$. Of first importance is to verify whether the procedure from section \ref{sec:proc} gives us spectral functions of fermions described by massive Dirac theory. It is important to note that both the holographic and the semiholographic spectral functions must contain a gap. This is because using the procedure we apply, the spectrum obtained in semiholography should correspond to strongly coupled gapped Dirac fermions. The self-energy, i.e., the holographic Green's function, then contains the effect of strong interactions between these fermions. If the gap were caused by the parameter $M_0$ only, the spectral function would instead correspond to gapped fermions interacting with a strongly coupled gapless CFT, which is not what we are after in this work. We therefore start this section by verifying the appearance of a gap in the holographic spectral functions. 
\subsubsection{Holographic spectra} \label{sss:spechol}
\begin{figure}[!t]
  \begin{subfigure}[b]{.46\linewidth}
	\centering 
	\includegraphics[clip=true, scale=.5]{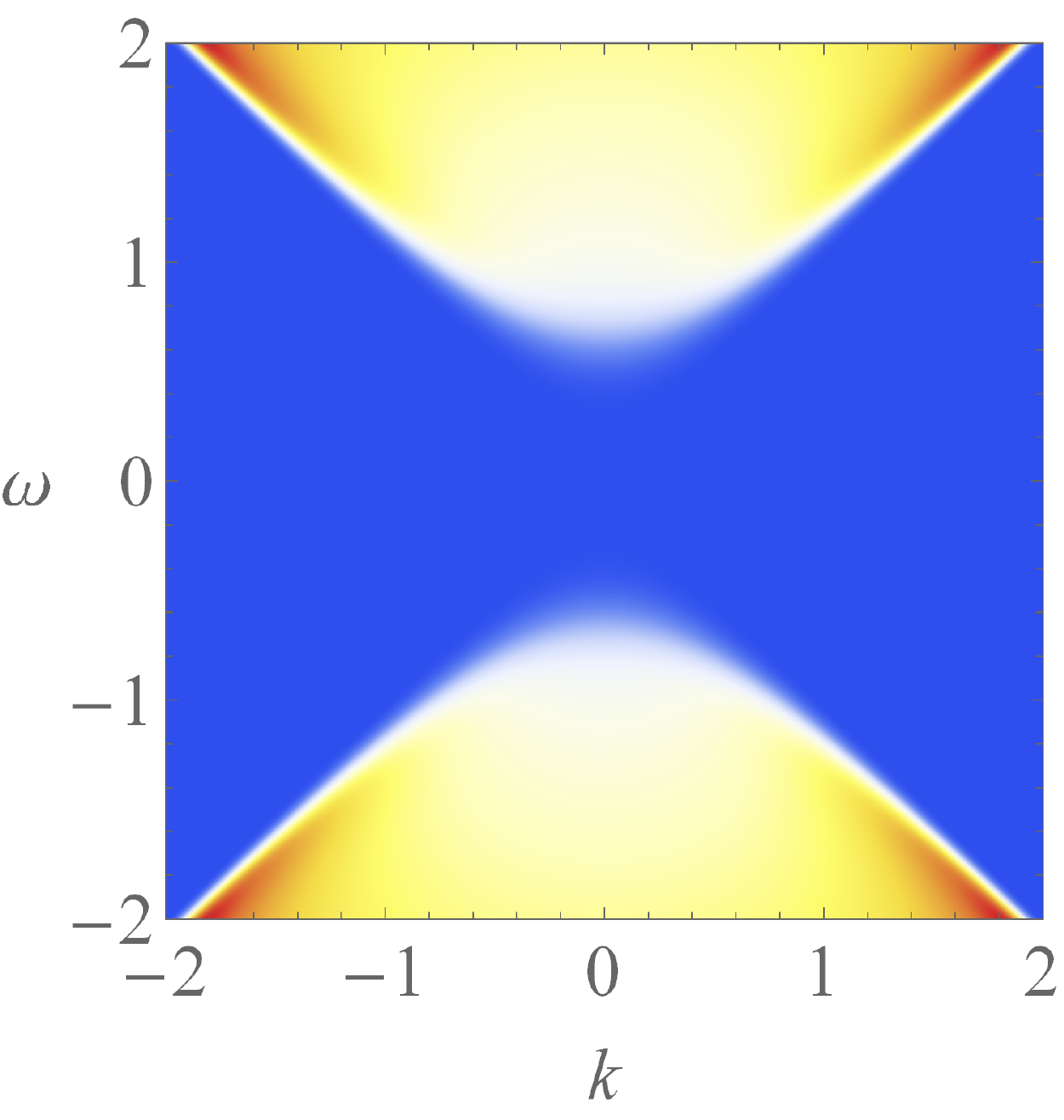}
		    \caption{}
		\label{fig:SpFhol}
  \end{subfigure}
  \begin{subfigure}[b]{.54\linewidth}
	\includegraphics[clip=true, scale=.5]{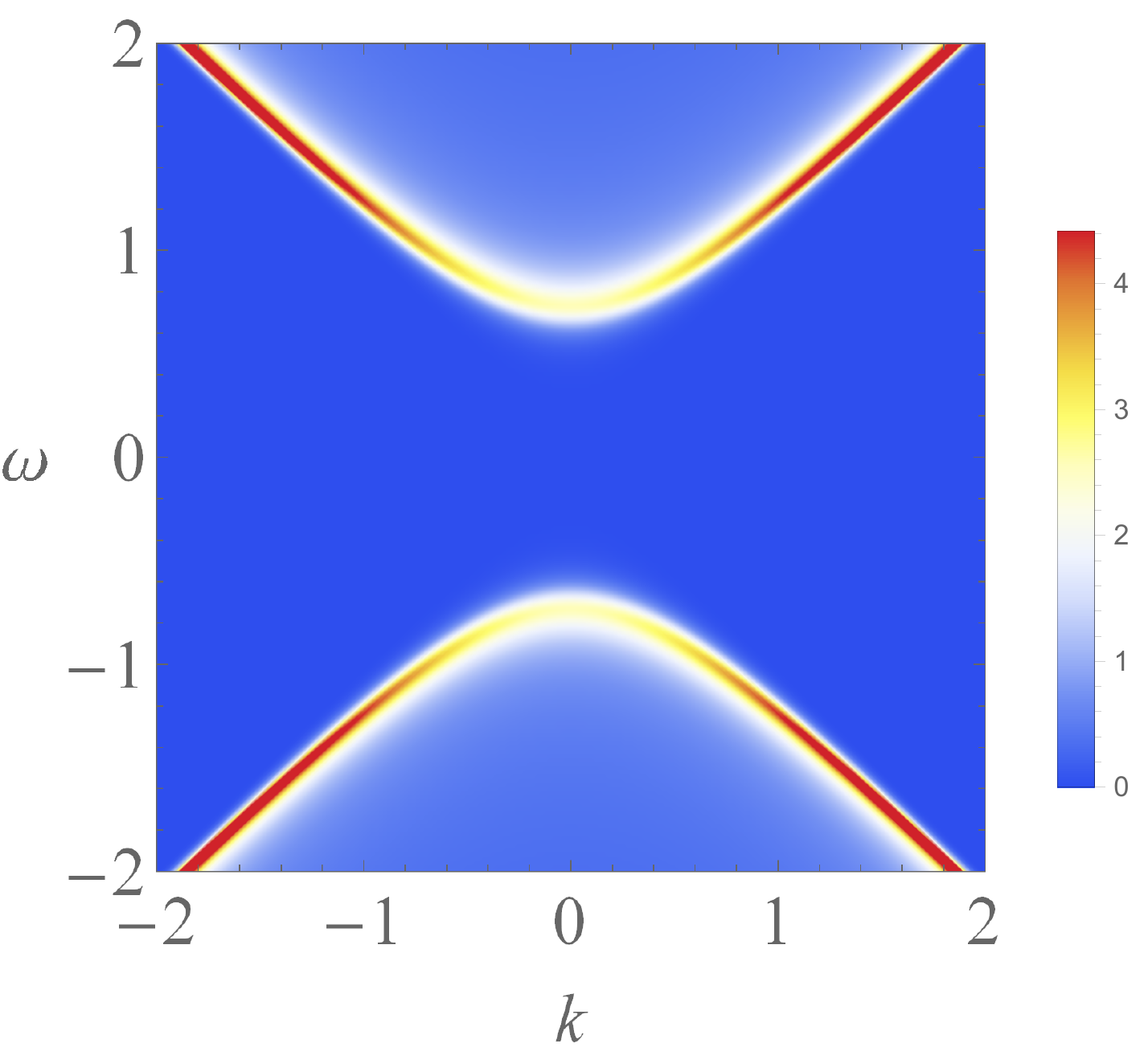}	
		 \caption{}
		\label{fig:SpFAQ}
  \end{subfigure}
\caption{\label{fig:holspec} (a) The holographic spectral function. (b) The holographic spectral function in alternative quantization. In both (a) and (b), $\lambda=1$. The legend on the right holds for both figures. Here, and in all the following plots, all quantities are made (scale) dimensionless by dividing by the appropriate power of $M_0$, and we choose $q=1$, $M=1/4$, $T=1/100$ unless stated otherwise.}
\end{figure}
\begin{figure}[!t]
  \begin{subfigure}[b]{.5\linewidth}
	\centering 
	\includegraphics[clip=true, scale=.5]{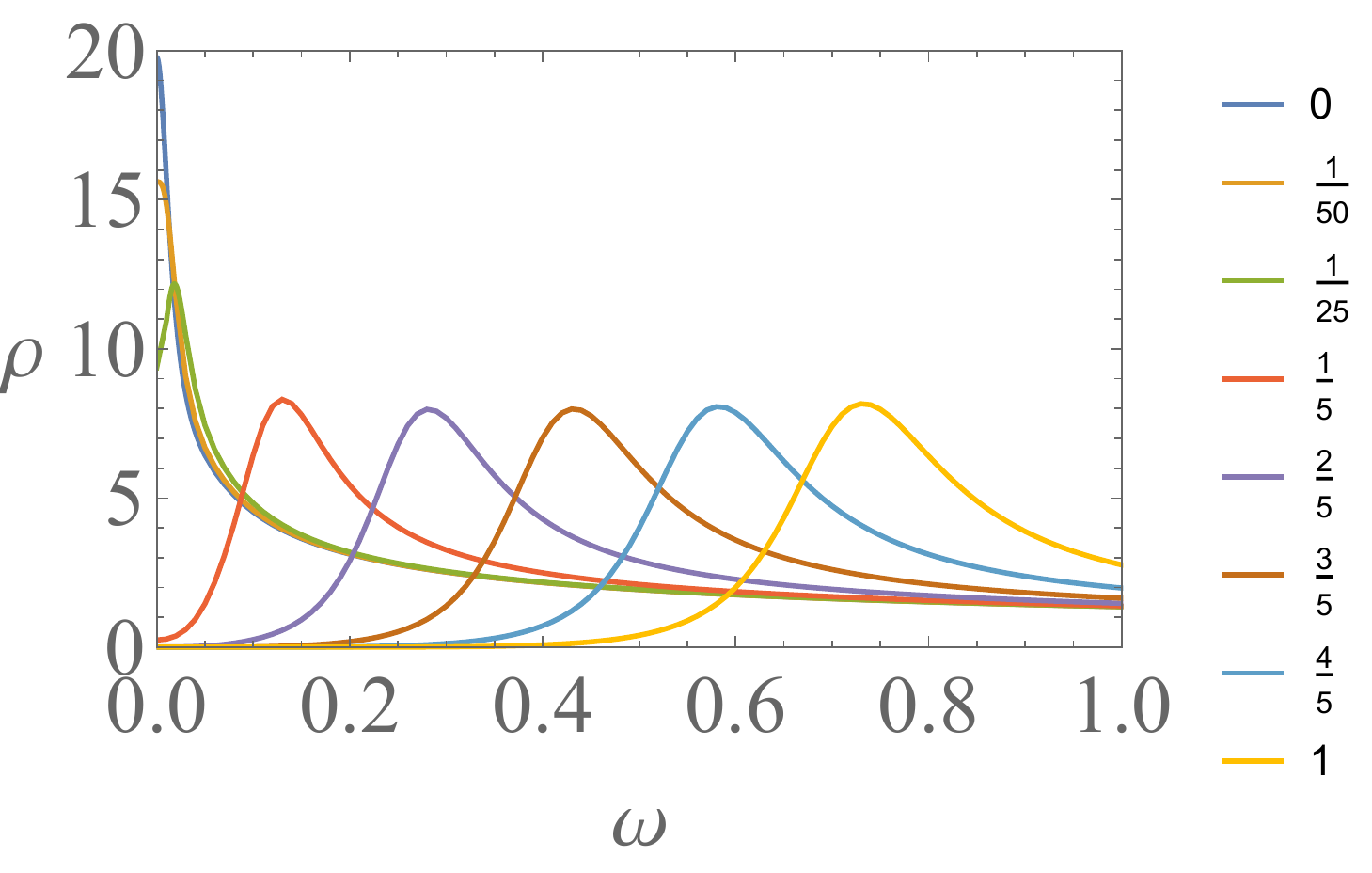}
		    \caption{}
		\label{fig:ImSiggc}
  \end{subfigure}
  \begin{subfigure}[b]{.5\linewidth}
	\includegraphics[clip=true, scale=.39]{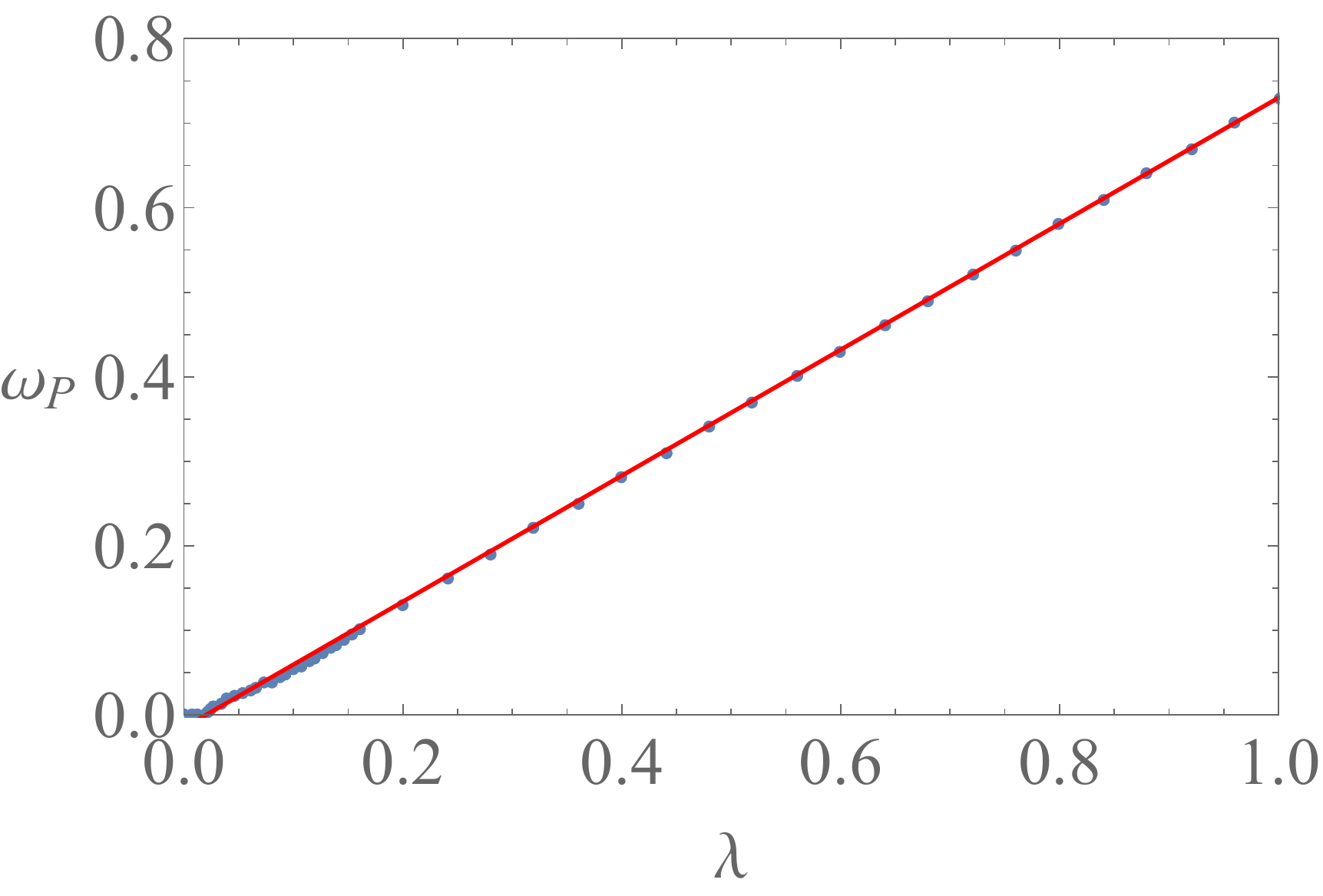}	
		 \caption{}
		\label{fig:wpeak}
  \end{subfigure}
\caption{\label{fig:gcdep} (a) The holographic spectral function in alternative quantization at zero momentum. The legend shows the used values of $\lambda$. (b) The dependence of the peak position $\omega_P$ on $\lambda$.}
\end{figure}
In figure \ref{fig:holspec} we show the holographic spectral functions for $\lambda=1$, in both standard and alternative quantization. These contain a gap as desired.\footnote{Note that this is not a hard gap.} To obtain this gap it is imperative that $\lambda$ is nonzero, since this parameter describes the coupling between chiral components. As a consequence, when $\lambda=0$ we expect no gap in the self-energy, and neither do we expect a peak at nonzero $\omega$ in alternative quantization. This is indeed the case, as is shown in figure \ref{fig:gcdep}, where we study the dependence on $\lambda$ of the peak appearing in the holographic spectral function in alternative quantization. We observe that a peak at nonzero $\omega$ appears for values of $\lambda$ higher than a nonzero lower bound. Furthermore, for larger values of $\lambda$, the position of the peak grows approximately linearly with $\lambda$, whereas the height remains almost constant. We expect these results by noting that \eqqref{eq:Xieq} only depends on the combination $\lambda\phi$, rather than $\lambda$ and $\phi$ seperately. Therefore, asymptotically the relevant scale is $\lambda M_0$ rather than $M_0$.\footnote{In holography the relevant scale is actually the source $\phi_s$ rather than $M_0$, but as stated before their ratio is fixed in this work.} In the low-temperature regime, $\lambda M_0$ is then the only dimensionful scale left and we therefore expect the peak to be proportional to $\lambda M_0$. This also explains the discrepancy observed in figure \ref{fig:wpeak} at low $\lambda$, since here the scale $T/\lambda M_0$ becomes large. We have indeed observed that a peak appears for smaller values of $\lambda$ as well when lowering the temperature further. However, the initial conditions corresponding to \eqqref{eq:redXieqs} depend on $\lambda$ but not on $M_0$. It is therefore not completely obvious to us that the position of the peak should grow linearly with $\lambda$, but the numerics show that this is indeed the case. Finally, notice that due to the width of the peak, a large enough value of $\lambda$ is needed before the gap appears. This spread is not solely due to the nonzero temperature, which we have checked numerically by calculating the same spectral functions at lower temperatures and not observing a decrease in the width. As a consequence the observed peak cannot correspond to a long-lived quasiparticle, which we indeed would not expect from a holographic spectral function describing unparticles in a mass-deformed conformal field theory.
\subsubsection{Conductivity}
Having shown that a gap is introduced in the holographic spectral functions, it is interesting to see if the CFT now indeed behaves as an insulator. We can check whether this is the case by calculating the conductivity of the CFT. In order to do so, we introduce fluctuations of the gauge field component $\delta A_x(x_\mu)=\delta a_x(r)e^{-i\omega t}$ to the theory. These fluctuations are not coupled to fluctuations of the other fields, even though the background has a nontrivial scalar profile. In particular, in contrast to the holographic superconductor model, these gauge fluctuations are not coupled to the scalar fluctuations $\delta\phi$ because the scalar field is uncharged. Moreover, the metric fluctuations $\delta g_{tx}$ are not sourced because we are still considering the undoped case. The fluctuations $\delta a_x$ then satisfy the equation of motion\footnote{In \eqqref{eq:condeq} and \eqqref{eq:axasymp}, we have not scaled dimensionful quantities like $\omega$ and $\delta a_{x(1)}$ by the mass $M_0$.}
\begin{equation} \label{eq:condeq}
	\delta a_x'' + \left( \frac{f'}{f} - \frac{\chi'}{2} + \frac{1}{r} \right) \delta a_x' + \frac{e^\chi \omega^2}{f^2} \delta a_x = 0.
\end{equation}
This equation has the asymptotic solution 
\begin{equation} \label{eq:axasymp}
\delta a_x = \delta a_{x(0)} + \delta a_{x(1)} r^{-2} + \frac{\omega^2}{2}\delta a_{x(0)} r^{-2} \log(r) + \cdots .
\end{equation}
By analyzing the action up to second order in the fluctuations, we can then find that the conductivity is given by \cite{Horowitz1}
\begin{equation} \label{eq:conductivity}
  \sigma(\omega) = 2\frac{\delta a_{x(1)}}{i \omega \delta a_{x(0)}} - \frac{\omega}{2i},
\end{equation}
where the coefficients $\delta a_{x(0)}$ and $\delta a_{x(1)}$ are found by solving \eqqref{eq:condeq} with infalling boundary conditions at the horizon.
\begin{figure}[!t]
  \begin{subfigure}[b]{.46\linewidth}
	\centering 
	\includegraphics[clip=true, scale=.45]{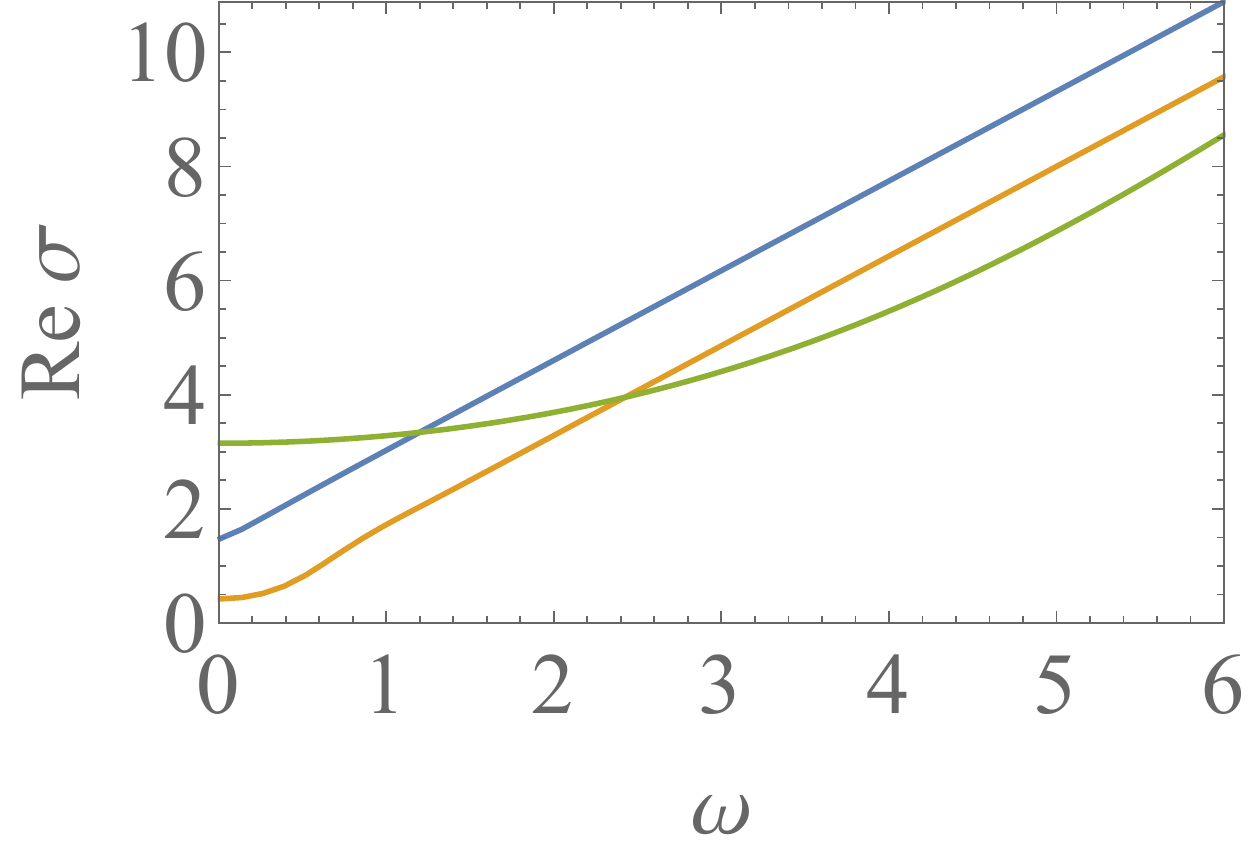}
		    \caption{}
		\label{fig:recond}
  \end{subfigure}
  \begin{subfigure}[b]{.54\linewidth}
	\includegraphics[clip=true, scale=.45]{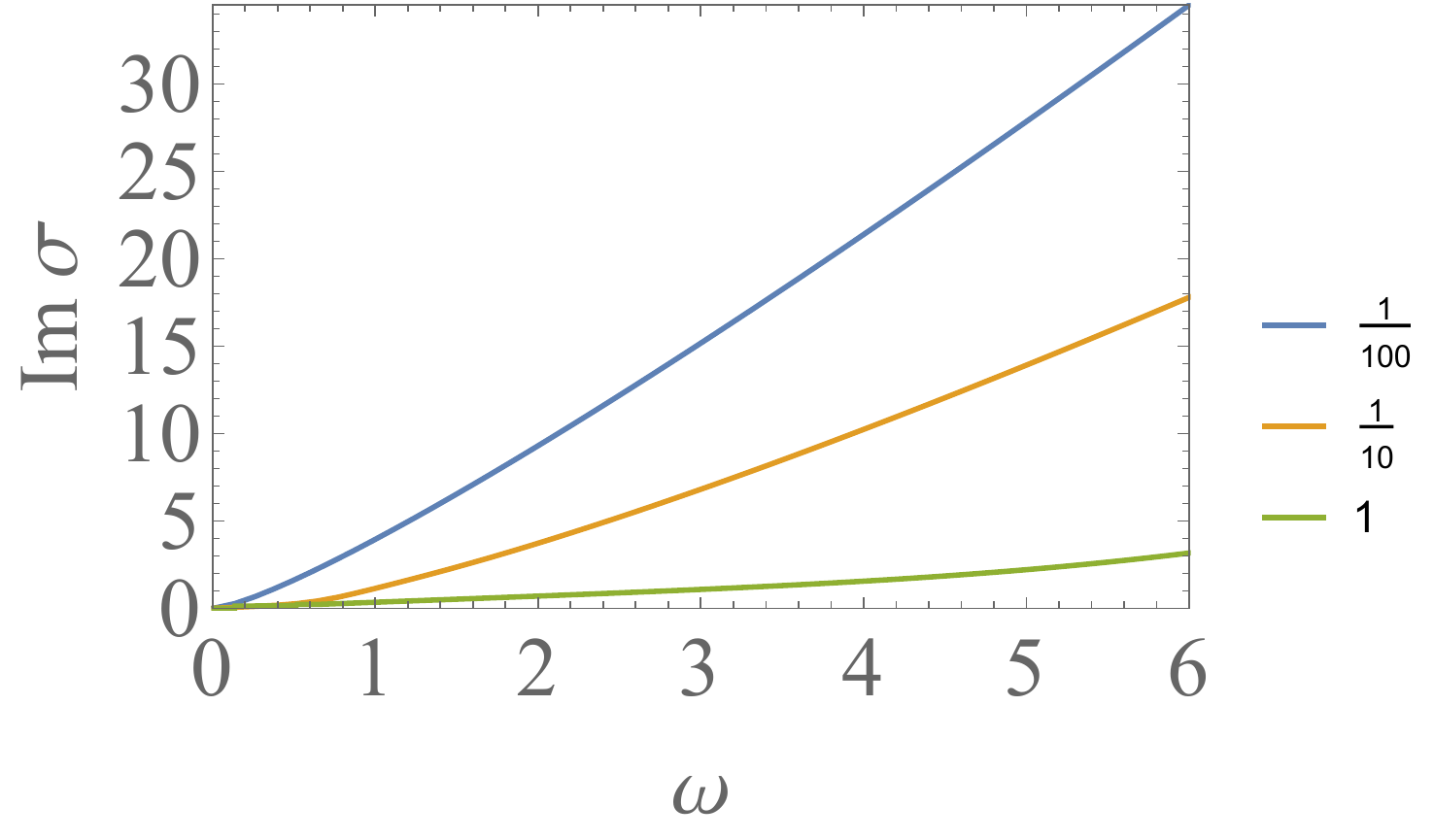}	
		 \caption{}
		\label{fig:imcond}
  \end{subfigure}
\caption{\label{fig:cond} The (a) real and (b) imaginary part the conductivity $\sigma$ of the CFT. The legend shows the value of the temperature for both figures.}
\end{figure}

The results shown in figure \ref{fig:cond} show that the conductivity does not behave as an insulator. For high temperatures the DC conductivity $\sigma(0)$ is linear in $T$, as we know from the conductivity in a Schwarzschild background \cite{Kovtun2}. For low temperatures, where the mass scale dominates, this linearity breaks down as expected. However, the mass scale does not induce a gap in the CFT conductivity. A possible explanation for this is the presence of other degrees of freedom in the CFT that are not gapped out by the mass deformation introduced in the model. A similar result was found in refs. \cite{Landsteiner1,Plantz1}. We could have anticipated this result as well from the fact that the calculation is independent of the coupling $\lambda$, which as we saw in the previous section generates the gap in the spectral functions. We expect however that the fermionic contribution to the conductivity, which can be calculated using the semiholographic fermionic Green's function using the approach explained in refs. \cite{Jacobs1,Jacobs2,Jacobs3}, does contain a gap and describes an insulator. 
\subsubsection{Semiholographic spectra}
The holographic spectra in section \ref{sss:spechol} show that the self-energy of the semiholographic Green's function contains a gap. Therefore, the gap in the semiholographic spectra is caused by both the bare mass $M_0$ and the gap in the self-energy. Note that here we are assuming a nonzero coupling $\lambda$. Otherwise, the Dirac field $\Psi$ would correspond to two independent Weyl fermions. Another way of seeing this is by noting from \eqqref{eq:redXieqs} that $\Xi_c=0$ when either $\lambda=0$ or $\phi=0$. 
\begin{figure}[!t]
\centering
\includegraphics[width=.65\textwidth,clip]{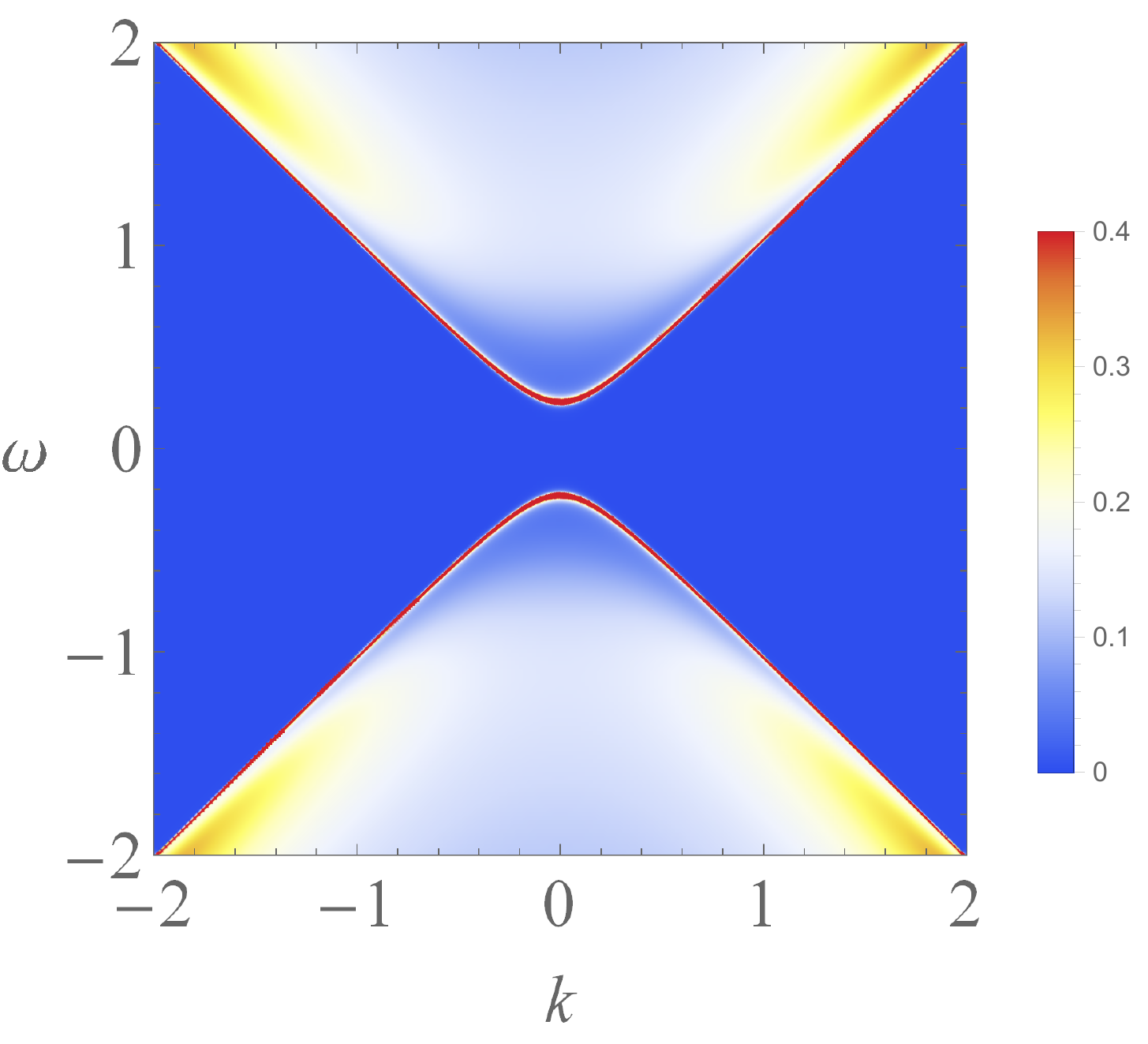}
\caption{\label{fig:SpFex} The undoped spectral function of the elementary fermion for $\lambda=1$ and $g=1$.}
\end{figure}
\begin{figure}[!t]
  \begin{subfigure}[b]{.5\linewidth}
	\centering 
	\includegraphics[clip=true, scale=.42]{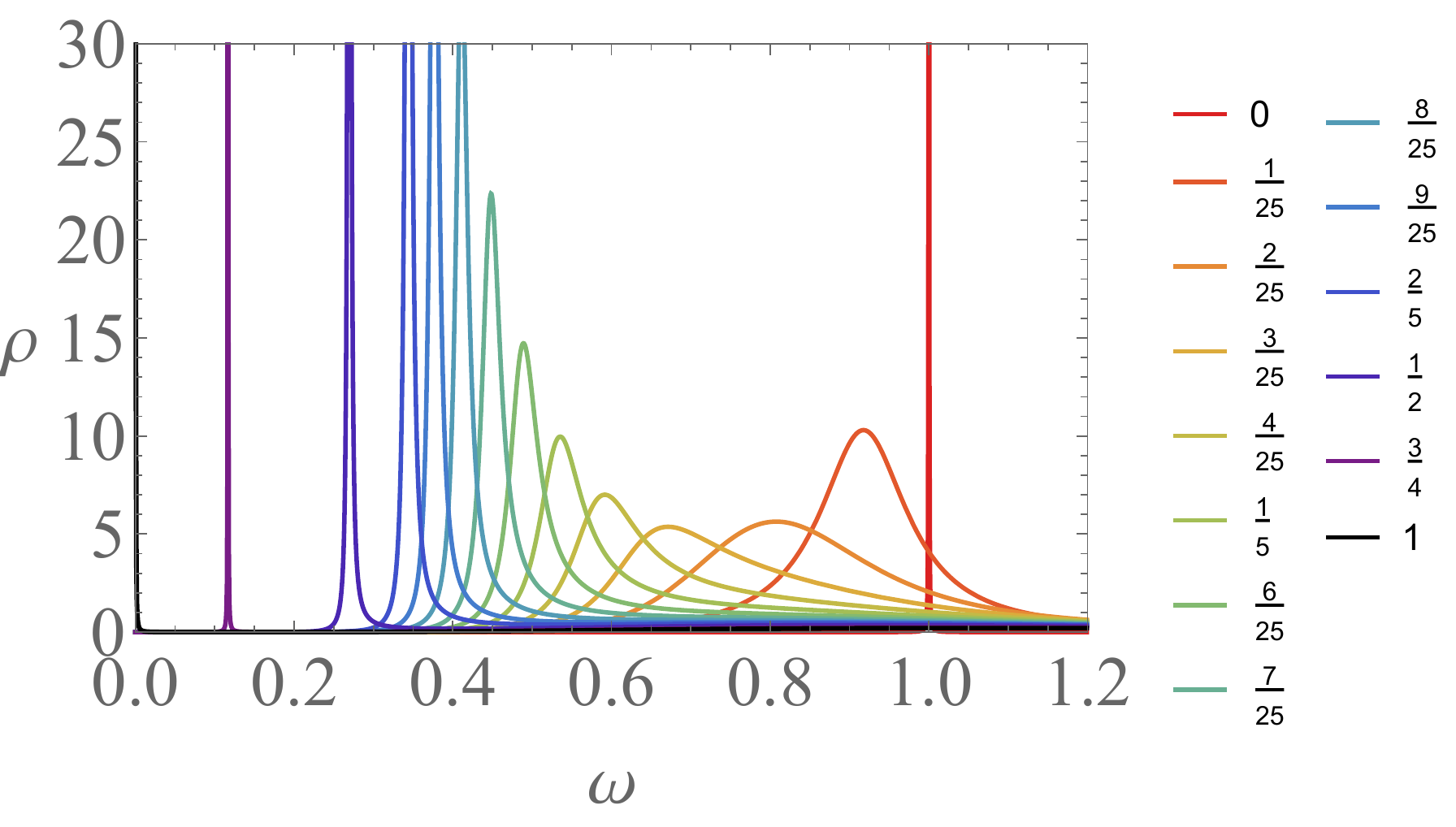}
		    \caption{$\lambda=1$}
		\label{fig:belowc1}
  \end{subfigure}
  \begin{subfigure}[b]{.5\linewidth}
	\includegraphics[clip=true, scale=.42]{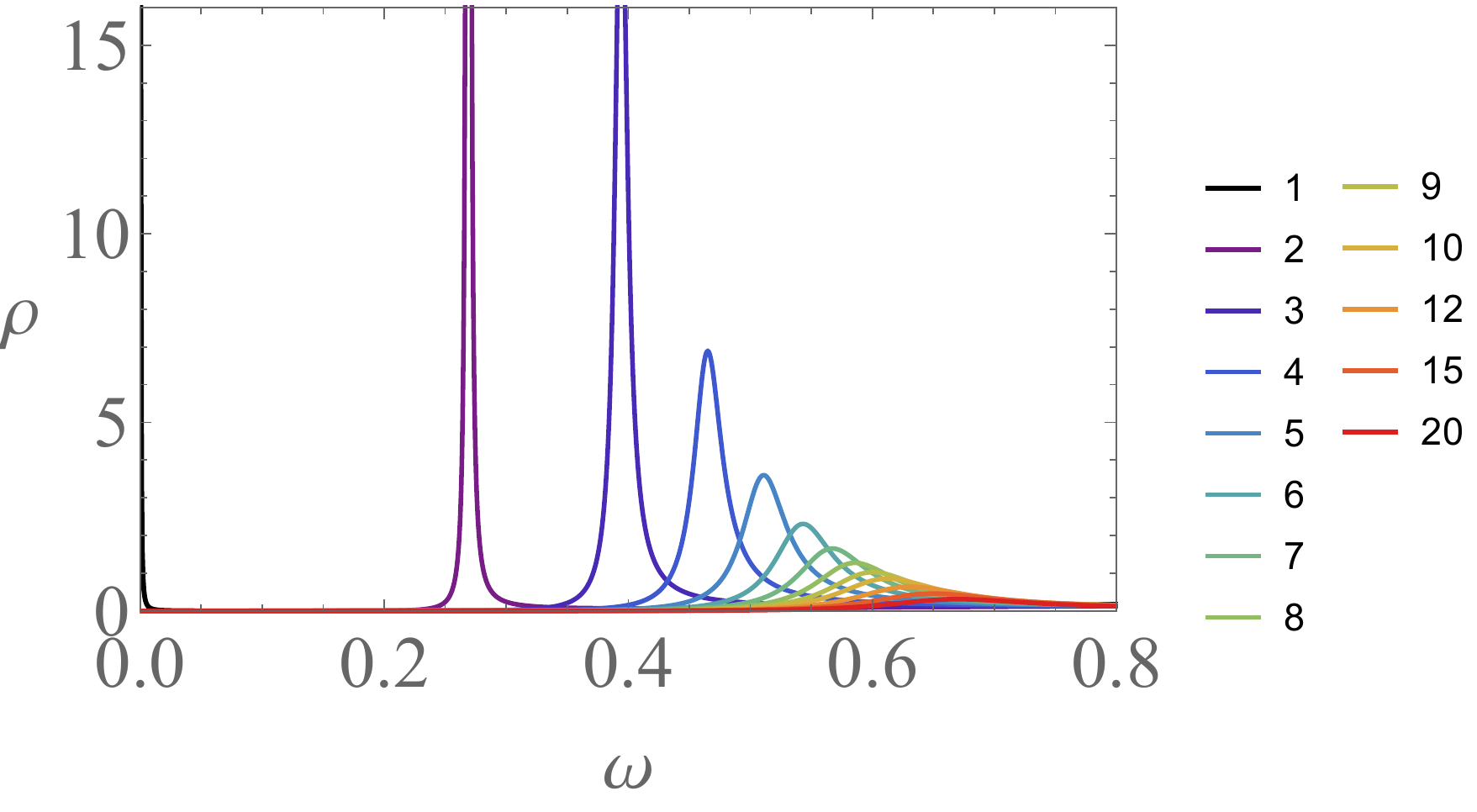}	
		 \caption{$\lambda=1$}
		\label{fig:abovec1}
  \end{subfigure}
\\ \\
  \begin{subfigure}[b]{.5\linewidth}
	\centering 
	\includegraphics[clip=true, scale=.47]{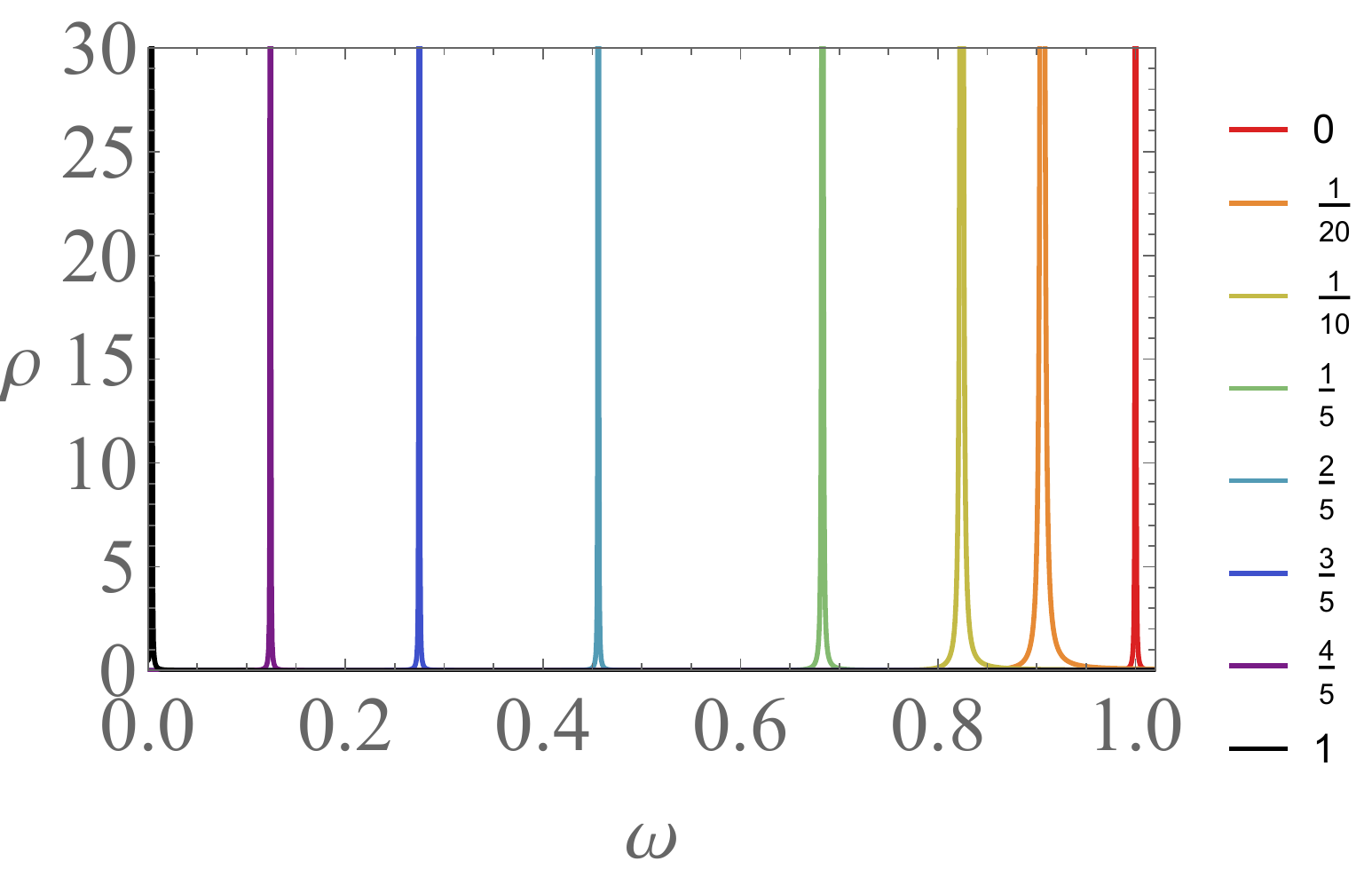}
		    \caption{$\lambda=2$}
		\label{fig:belowc2}
  \end{subfigure}
  \begin{subfigure}[b]{.5\linewidth}
	\includegraphics[clip=true, scale=.45]{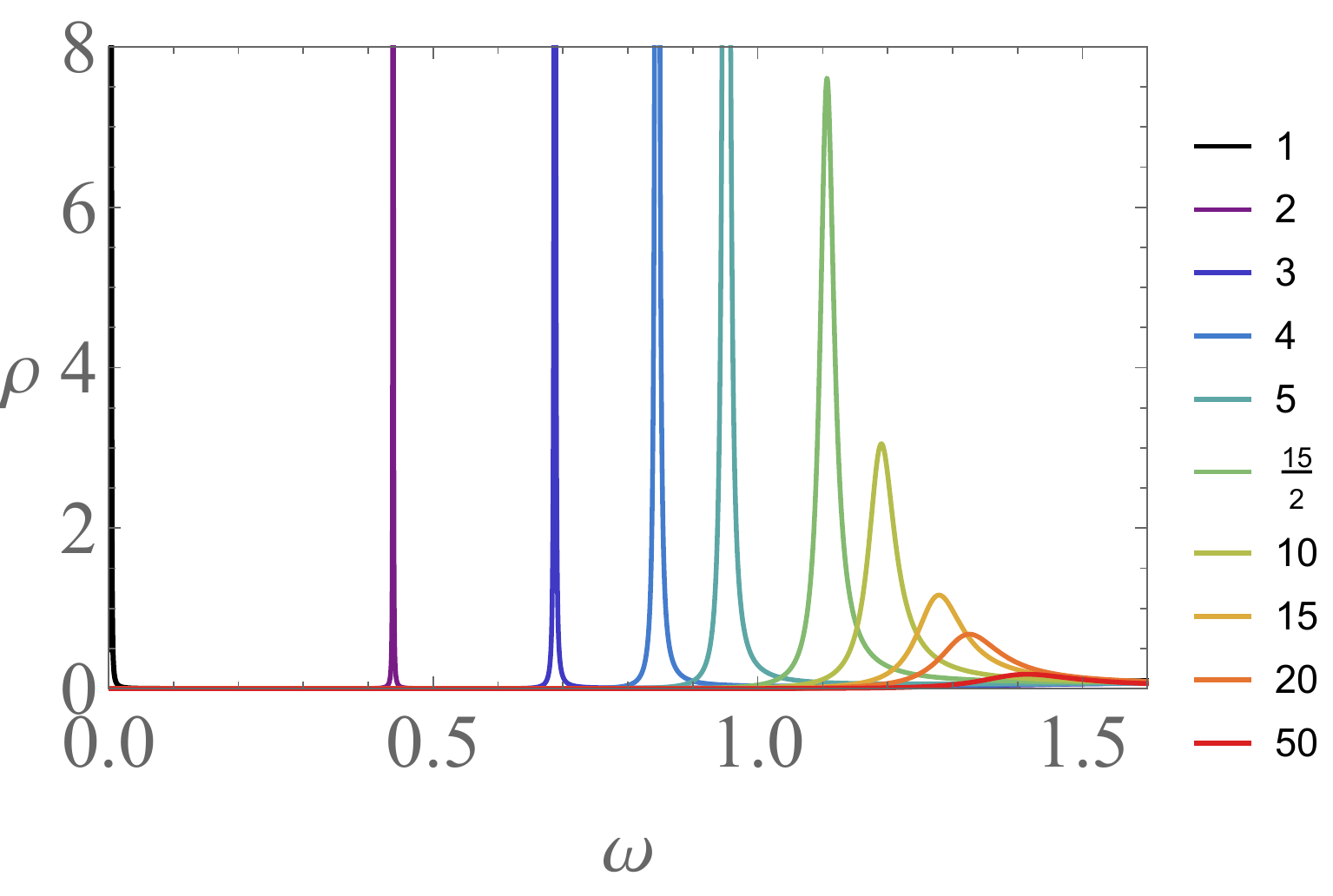}	
		 \caption{$\lambda=2$}
		\label{fig:abovec2}
  \end{subfigure}
\caption{\label{fig:semiholg} The semiholographic spectral function at zero momentum for several values of $g$. The legends show the value of $g/g_c$. In (a) and (b), $\lambda=1$ and $g_c=0.56$. In (c) and (d), $\lambda=2$ and $g_c=0.39$. These spectral functions are symmetric in $\omega$ due to particle-hole symmetry. In (c) the peaks are all very sharp because they are at frequencies inside the gap of the self-energy for $\lambda=2$.}
\end{figure}

In figure \ref{fig:SpFex} we show the semiholographic spectral function for $\lambda=1$ and $g=1$, which also contains a gap.  For the specific set of parameters chosen there, we see that the mass is renormalized to a value smaller than the bare mass. From our previous analysis of figure \ref{fig:wpeak}, which gives the values of the renormalized mass for large $g$, we know that this is not always the case. This is again evident in figure \ref{fig:semiholg}, where we study the dependence of the spectral functions at zero momentum on the coupling constants $g$ and $\lambda$. The values of $\lambda$ in the figures are large enough such that the self-energy has a gap. Clearly, for $g=0$ the spectral function resembles that of a free Dirac fermion with mass $M_0$. Upon increasing $g$, for both values of $\lambda$  in the figures the mass first renormalizes to a smaller value, but ultimately converges for large $g$ to the value shown in figure \ref{fig:wpeak} that is obtained in the holographic spectral functions, which depending on the value of $\lambda$ can be either smaller or bigger than the bare mass. A remarkable feature in both the cases shown is that there exists a value of $g$ at which the mass renormalizes to zero. This can be understood from the general form of the Green's function in \eqqref{eq:renGR}, where we see that the mass is renormalized with a value proportional to $g$. To be more precise, by studying the symmetries of the equations in \eqqref{eq:redXieqs}, we can write the Green's function for small values of $\omega$ and $k_3$ as 
\be
G_R^{-1}(k)=\begin{pmatrix} Z_0 \sigma\cdot k & iM_{\text{eff}}\\ -iM_{\text{eff}} & -Z_0\bar{\sigma}\cdot k \end{pmatrix}
\ee
where $Z_0$ and $M_{\text{eff}}$ are given by
\ba
Z_0&=1+g\partial_\omega\Xi_0 (k_\mu=0), \label{eq:Z}\\
M_{\text{eff}}&=M_0+i g\Xi_c(k_\mu=0).\label{eq:Meff}
\ea
Referring to appendix \ref{app:sym} for details, we note that $Z_0$ and $M_{\text{eff}}$ are real constants and $Z_0>0$. From this expression it follows that the effective mass changes sign when $g$ assumes the critical value  
\be \label{eq:gcrit}
g_c=\frac{iM_0}{\Xi_c(k_\mu=0)}.
\ee
\begin{figure}[!t]
\centering
	\includegraphics[clip=true,scale = 0.5]{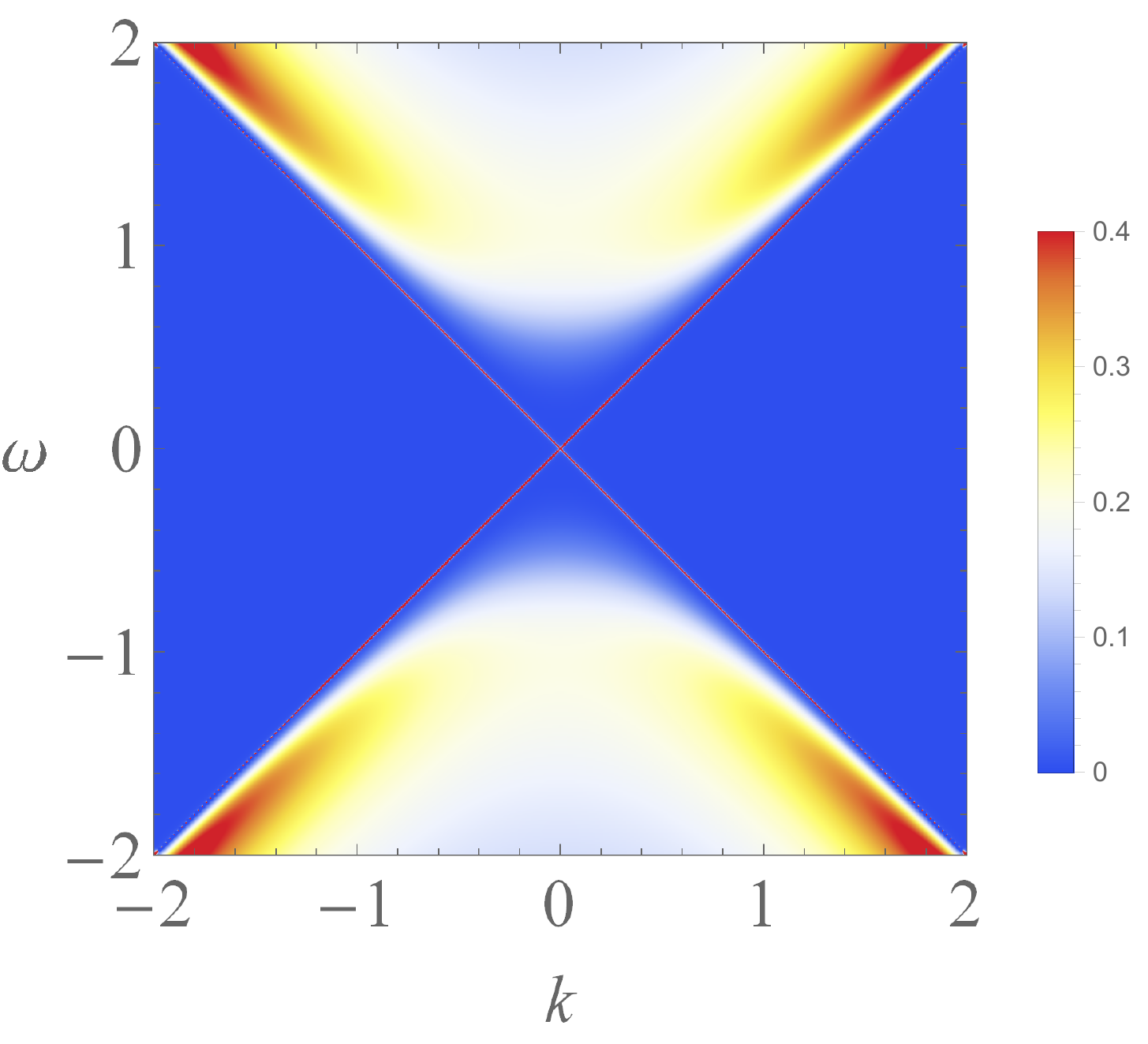}	
\caption{\label{fig:SpFQCP} The spectral function at $\lambda=1$ and $g=g_c=0.56$.}
\end{figure} \\
This signals a topological quantum phase transition similar to the one obtained by changing the sign of the mass in free Dirac theory\footnote{Strictly speaking we cannot see the quantum phase transition, since the temperature in our numerical computation is never exactly equal to zero.} \cite{Schnyder1,Volovik1,Volovik2}. This can for example be seen by defining a winding number as in ref. \cite{Volovik1}, which changes when inverting the sign of the Dirac mass.\footnote{Alternatively, this can be done by studying the behavior of the eigenspinor components of the Dirac Hamiltonian under a parity transformation.} We note that this transition is topological only when the symmetry protecting this winding number is not broken during the transition. Due to the symmetry of the Dirac equation it is not possible to determine which sign of the mass corresponds to a topologically trivial or nontrivial state. However, regardless of the initial sign of the mass of a state, we can say that to adiabatically transform this state into a state with a changed sign of the mass, i.e., a changed winding number, requires going through a gapless state, given that the protecting symmetry is respected during this transition. At the quantum critical point, $M_{\text{eff}}=0$ and the spectrum looks like that of a massless quasiparticle, as shown in figure \ref{fig:SpFQCP}. Since the dispersion of the peak now resides inside the gap, this time it does look like an infinitely long-lived quasiparticle. 

\begin{figure}[!t]
\centering
	\includegraphics[clip=true,scale = 0.53]{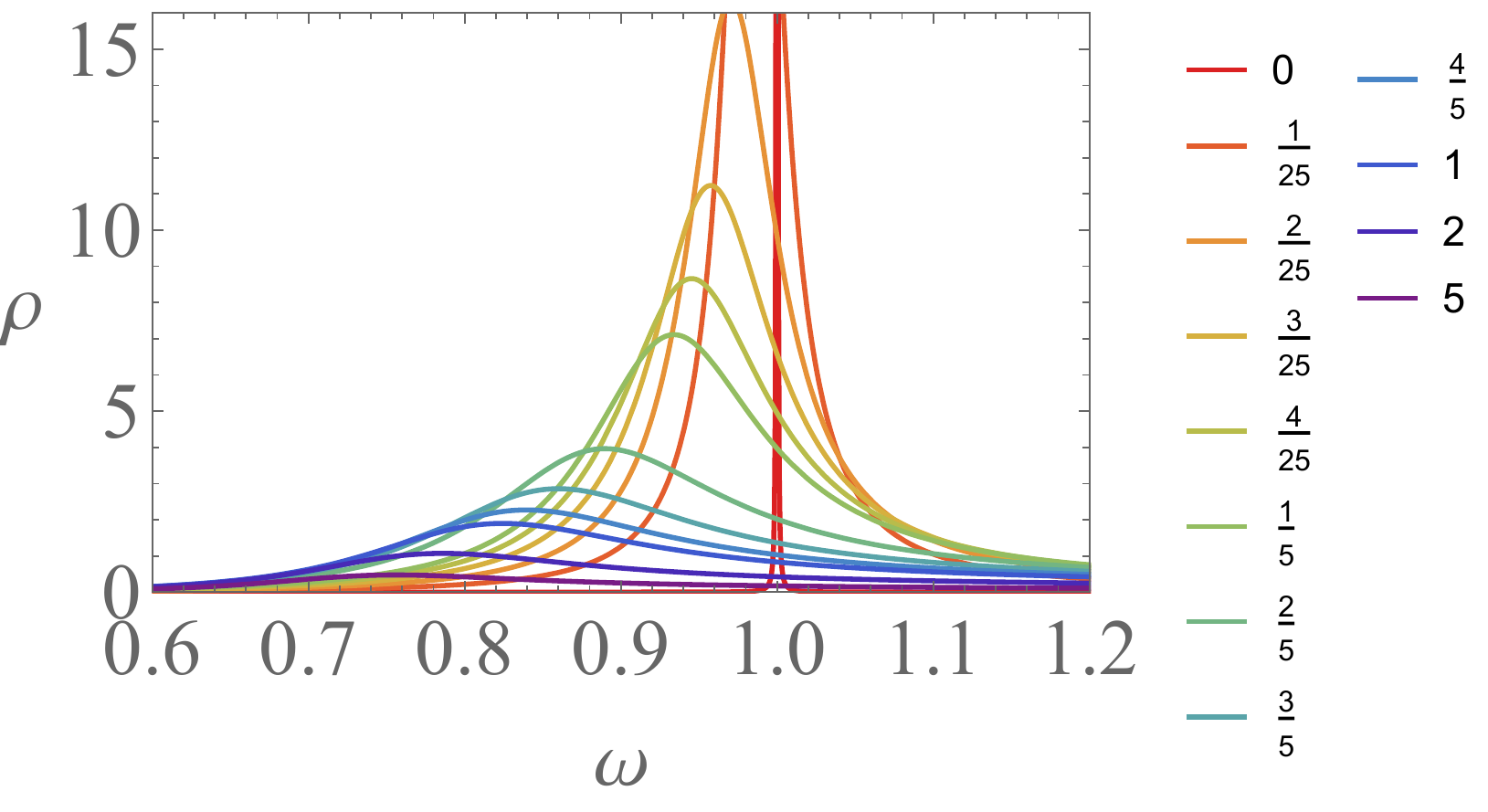}	
\caption{\label{fig:gcmin} The spectral function at $\lambda=-1$ for several values of $g$. The legend shows the value of the dimensionless coupling $g$.}
\end{figure}
What we have thus shown is that the introduction of the additional scale $g$ in semiholography induces a topological quantum phase transition. This scale is restricted to nonnegative values. Therefore, having obtained a solution for $\Xi$, it is possible to find a quantum critical solution by choosing $g$ as in \eqqref{eq:gcrit}, but only if $\text{Im}[\Xi_c(k_\mu=0)]>0$. It turns out that for $\lambda>0$ this is always the case, although $\lambda$ should be large enough to create a gap for \eqqref{eq:gcrit} to hold. Hence for positive $\lambda$, the Dirac fermion described by the holographic limit $g\rightarrow \infty$ will always be topologically distinct from the free fermion described by $g=0$. From the symmetry described in appendix \ref{app:sym} we can also immediately see that for $\lambda<0$ this is not the case, since then $\text{Im}[\Xi_c(k_\mu=0)]<0$. This can also be seen in figures \ref{fig:gcmin} and \ref{fig:SpFomegagc}. Moreover, from the symmetry described in appendix \ref{app:sym} that relates the self-energy corresponding to the bulk mass $M$ to the one for $-M$, it follows that this conclusion does not change when changing the sign of $M$. Specifically, this symmetry implies that when changing only the sign of $M$, the new value of $g_c$ is equal to the inverse of the old one. In fact, numerical analysis shows that $g_c$ is proportional to $\lambda^{-2M}$. Again, we can partially understand this by noting that in $\eqqref{eq:redXieqs}$ the asymptotically relevant scale is $\lambda M_0$ rather than $M_0$, so that at low temperatures and $k_\mu=0$ the asymptotic equation yields that $\Xi_c\propto (\lambda M_0)^{2M}$. However, solving the equation asymptotically yields an integration constant of which it is not entirely clear to us why its dependence on $\lambda$ is negligible.

\begin{figure}[!t]
\centering
	\includegraphics[clip=true,trim = 0cm 0cm 0cm 1cm, scale = 0.25]{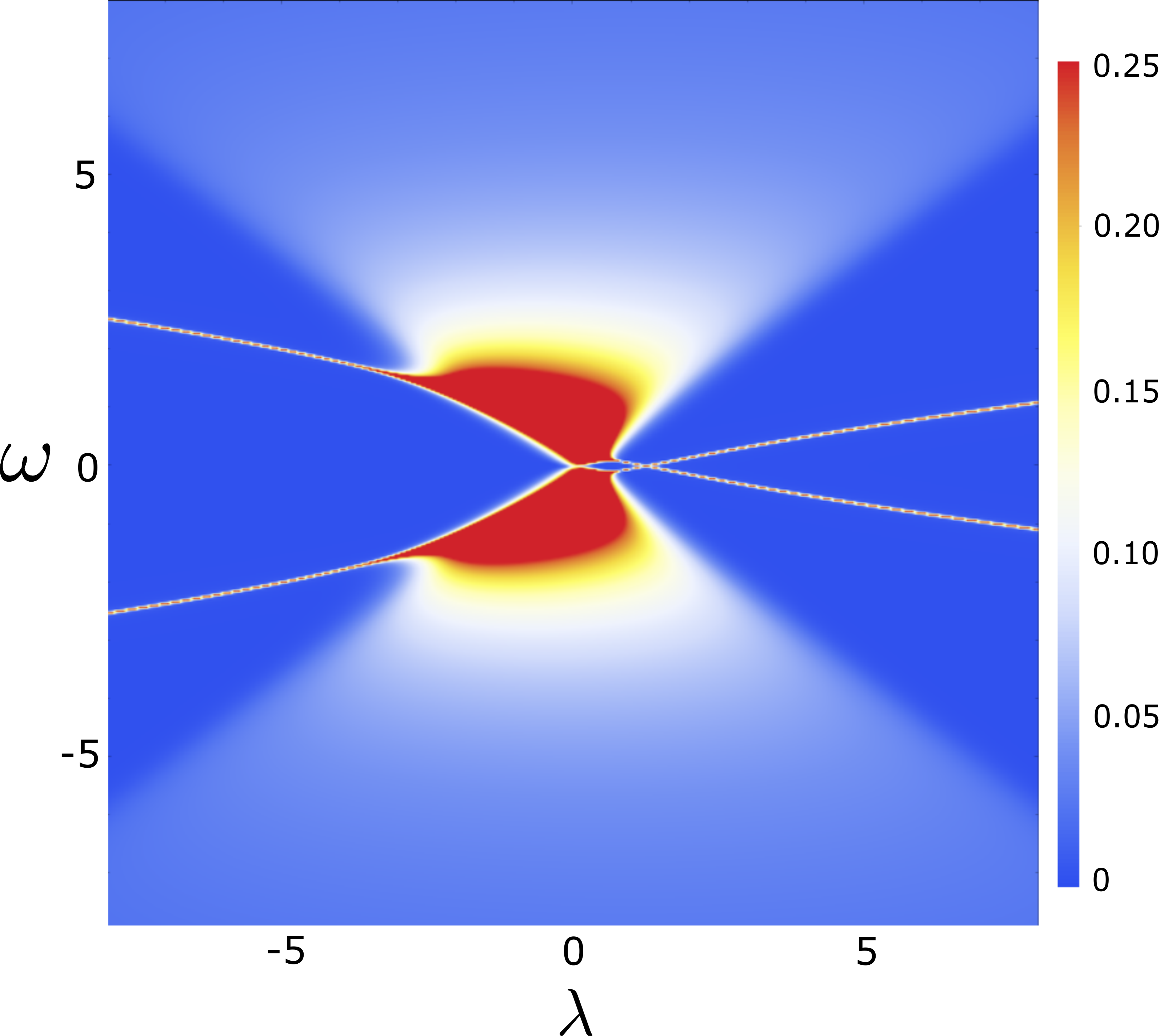}
\caption{\label{fig:SpFomegagc} The spectral function at $g=0.5$ and zero momentum as a function of $\lambda$. Here, the quantum critical point is clearly visible for $\lambda=1.25$, whereas there is no critical point for a negative value of $\lambda$.}
\end{figure}
All the spectral functions presented in this section satisfy the sum rule in \eqqref{eq:sumrule}, which we have verified numerically. However, in e.g. figures \ref{fig:SpFQCP} and \ref{fig:SpFex}, we see that for nonzero $g$ there is also spectral weight at frequencies higher than the position of the peak. This implies that due to interactions the spectral weight of the peak decreases. Numerically integrating over frequency reveals that for the case presented in figure \ref{fig:SpFex} the two red peaks carry less than half of the total spectral weight, which shows the significant effect of the interactions. The spread in the spectral weight continues to grow for higher $g$. In figure \ref{fig:SpFomegagc}, for large enough values of $|\lambda|$ we can clearly distinguish the spectral weight of the peaks due to the renormalized bare mass from the weight that originates from the self-energy. This self-energy contains a continuum rather than a peak, as we can see from figure \ref{fig:SpFhol}. Therefore, rather than an avoided crossing, we observe that the spectral weight merges into one broad peak for small values of $\lambda$. Moreover, in this figure we also see once again that a quantum phase transition occurs at a critical positive value of $\lambda$.
\subsection{Doped spectra}
We now turn to the case of nonzero chemical potential. Here we restrict ourselves to $\mu>0$, as the solution for $\mu<0$ then easily follows from particle-hole symmetry as described in appendix \ref{app:sym}. One trivial effect of the chemical potential is that the spectrum will appear shifted in frequency due to the use of grand-canonical energies. More interesting effects such as the formation of Fermi surfaces occur for large enough values of $\mu$. In this subsection we firstly study holographic spectral functions containing such Fermi surfaces. We then proceed with the semiholographic Green's functions with which we can also compute the corresponding momentum distribution functions. Finally, we study the dependence on the couplings $g$ and $\lambda$ of the characteristics of the theory near the Fermi surfaces.
\begin{figure}[!t]
  \begin{subfigure}[b]{.5\linewidth}
	\centering 
	\includegraphics[clip=true, scale=.424]{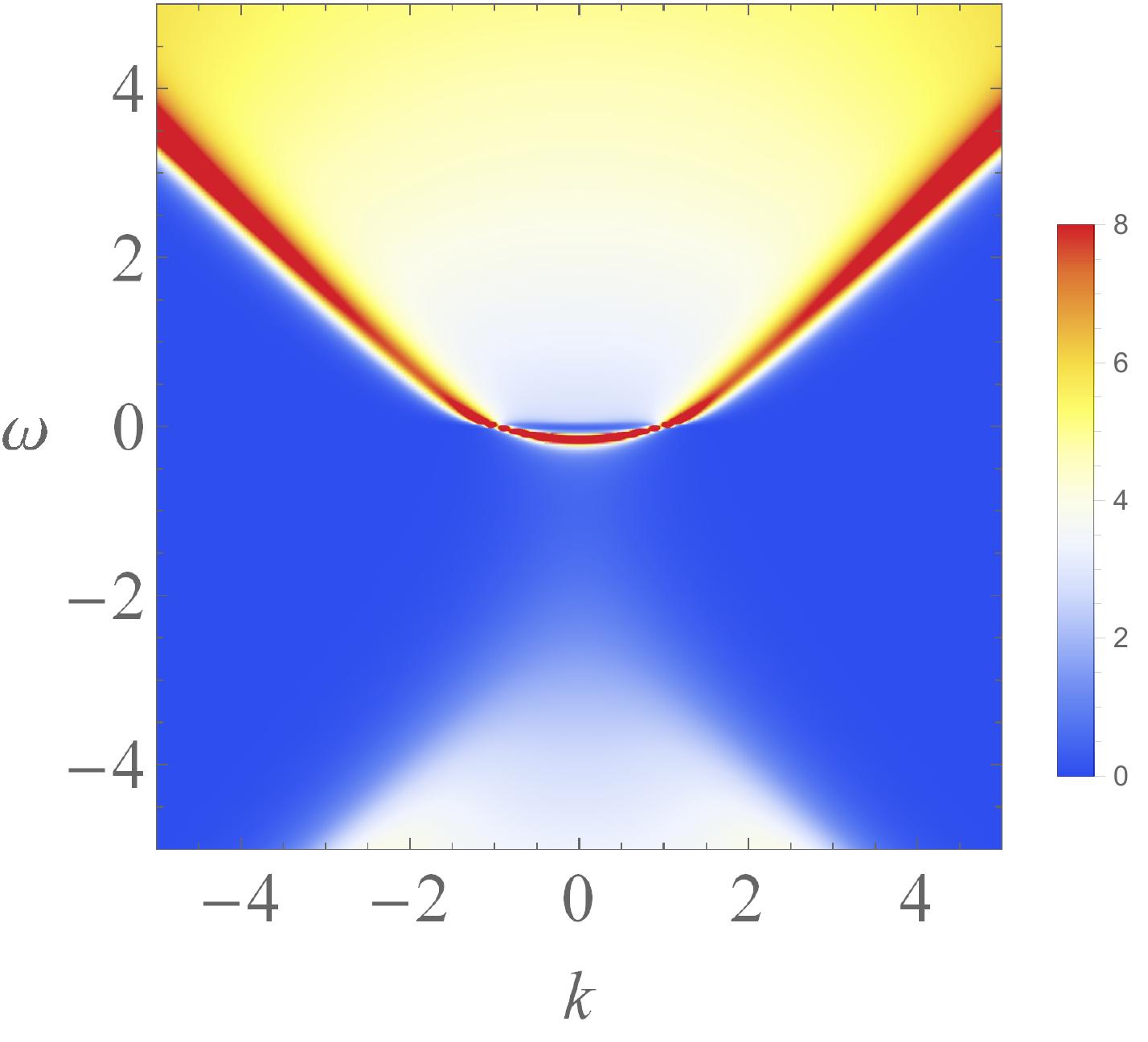}
		    \caption{}
		\label{fig:SpFmuholsq}
  \end{subfigure}
  \begin{subfigure}[b]{.5\linewidth}
	\includegraphics[clip=true, scale=.424]{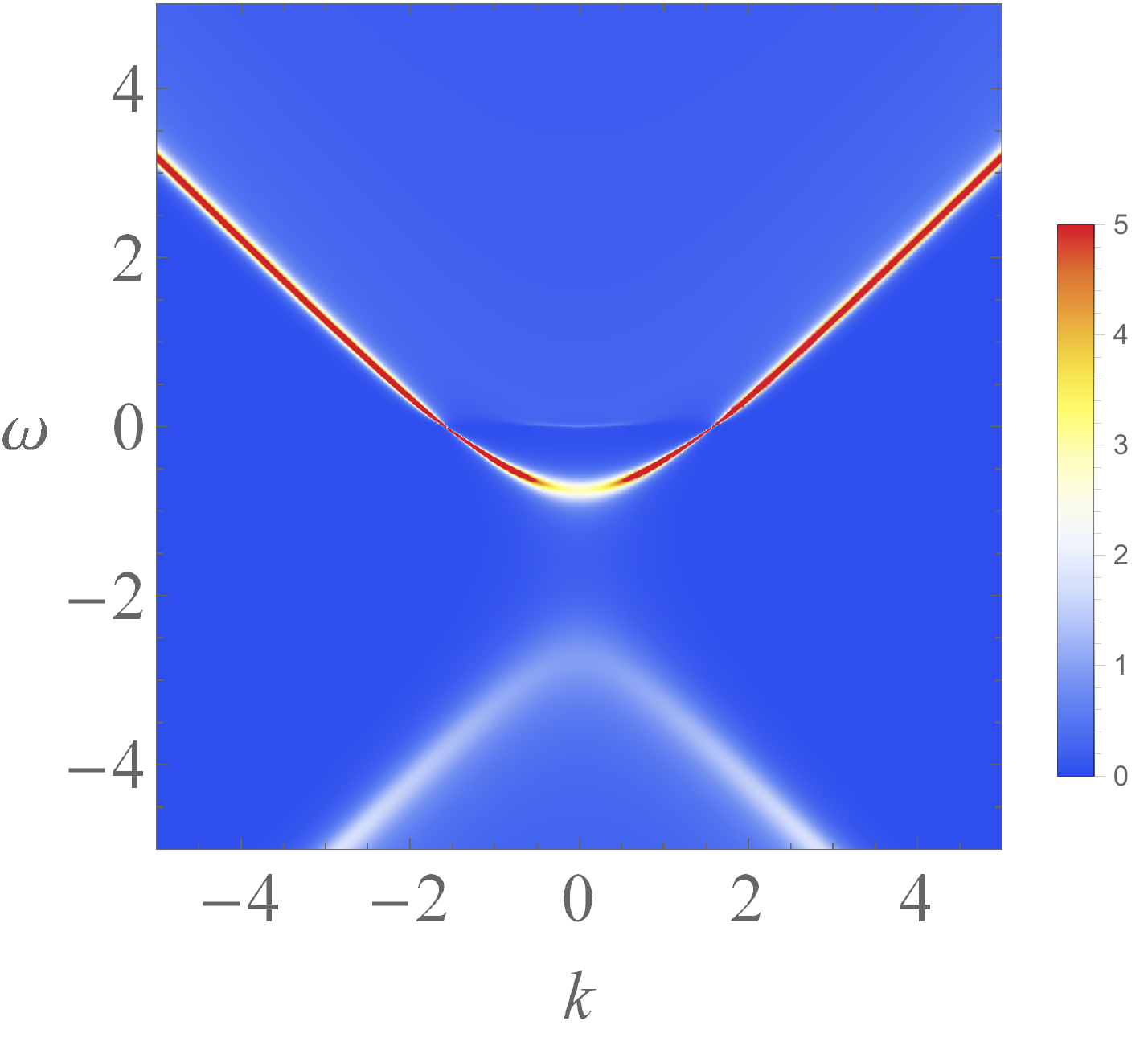}	
		 \caption{}
		\label{fig:SpFmuhol}
  \end{subfigure}
\caption{\label{fig:muspechol} The holographic spectral function for $\mu=2$ and $\lambda=1$ in (a) standard and (b) alternative quantization. The Fermi surface is most clearly visible in alternative quantization, where $k_F=1.58$.}
\end{figure}
\subsubsection{Formation of Fermi surfaces in the holographic spectra}
\begin{figure}[!t]
  \begin{subfigure}[b]{.5\linewidth}
	\centering 
	\includegraphics[clip=true, scale=.278]{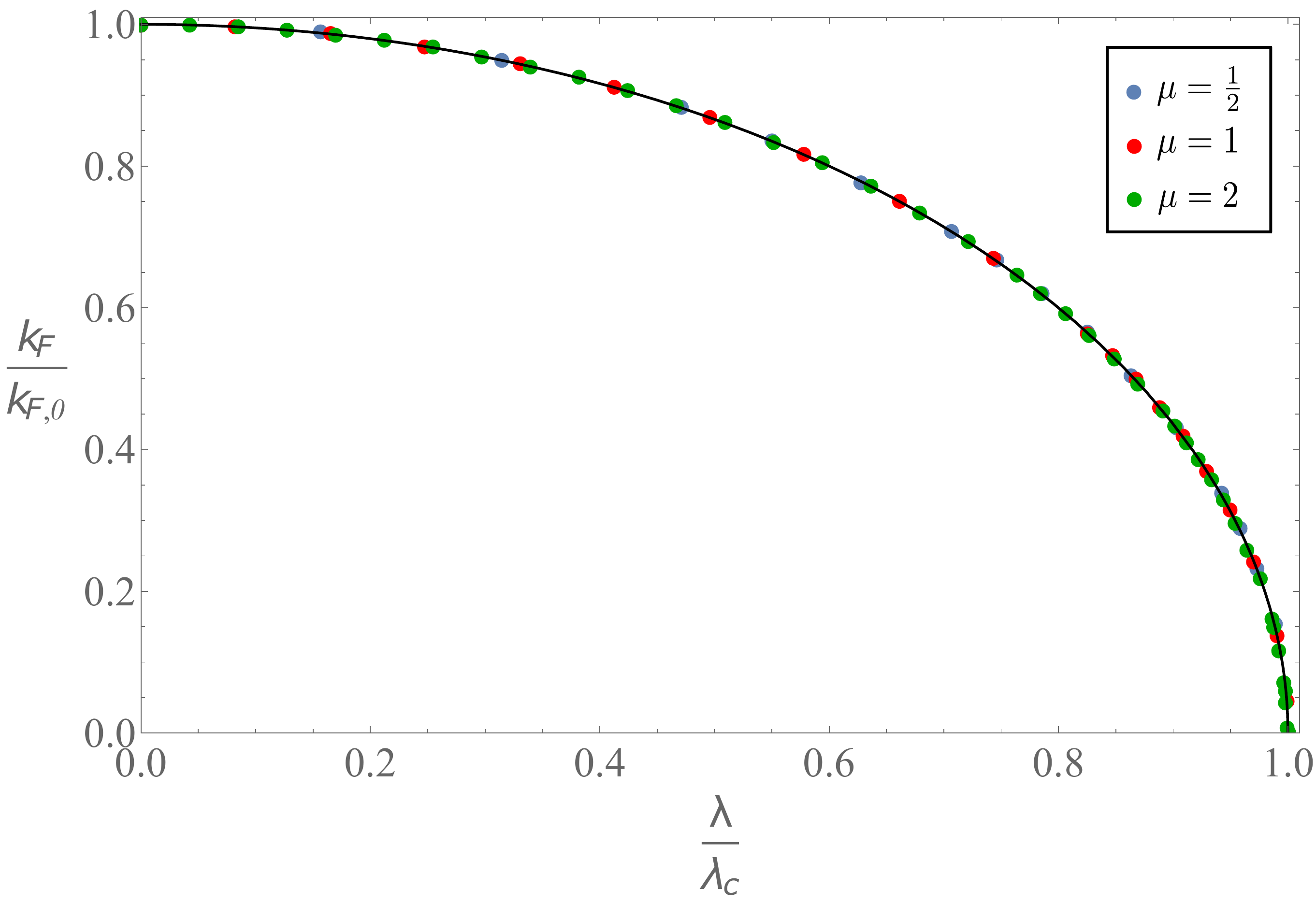}
		    \caption{}
		\label{fig:kFholsc}
  \end{subfigure}
  \begin{subfigure}[b]{.5\linewidth}
	\includegraphics[clip=true, scale=.453]{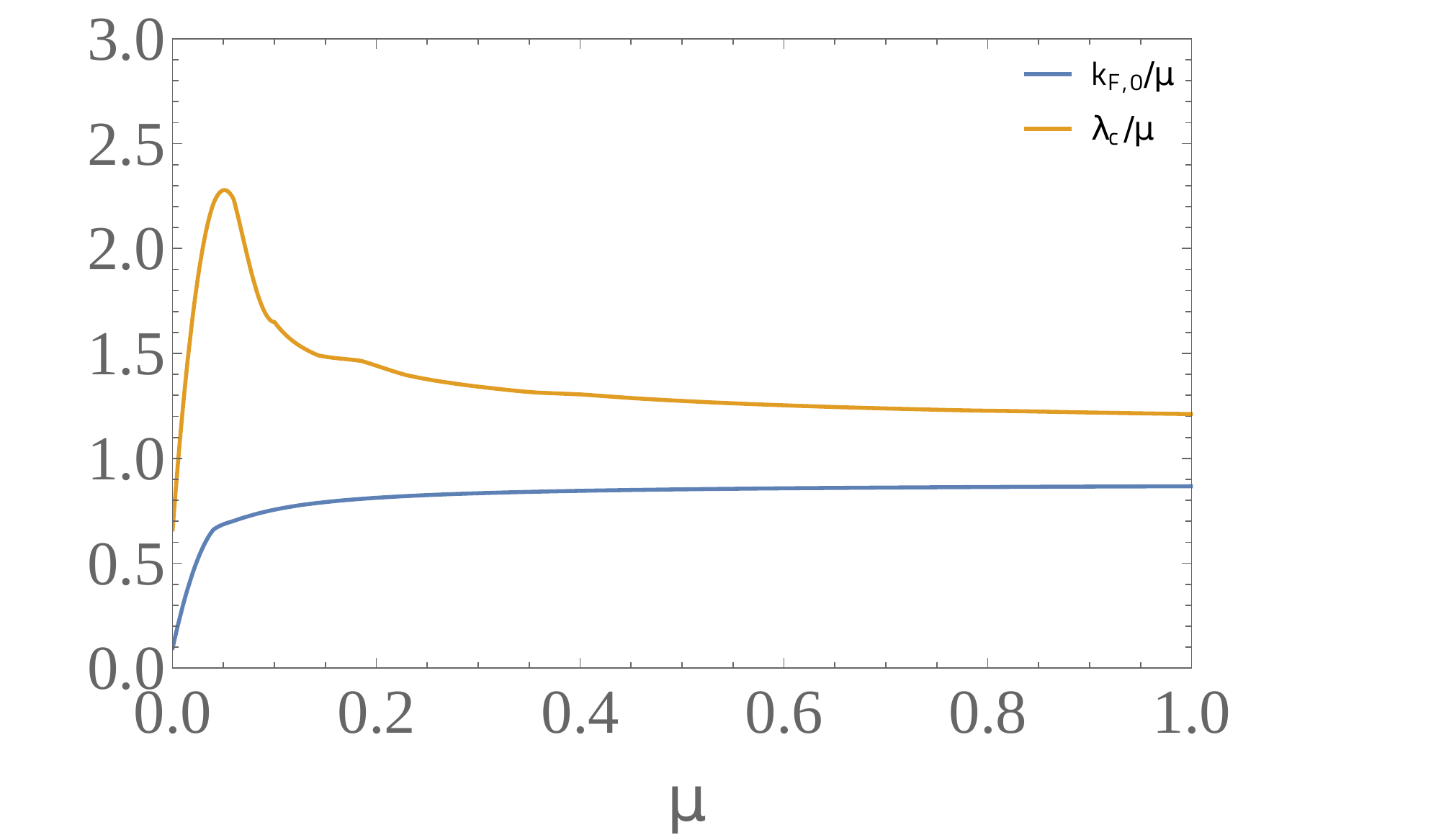}	
		 \caption{}
		\label{fig:kF0gc}
  \end{subfigure}
\caption{\label{fig:kFhol} (a) The Fermi momentum $k_F$ corresponding to the holographic Green's function in alternative quantization as a function of the coupling $\lambda$. The Fermi momentum is scaled with its value at $\lambda=0$ and the coupling $\lambda$ is scaled with the value $\lambda_c$ at which the Fermi surface vanishes. The black curve corresponds to \eqqref{eq:kFscale}. (b) The Fermi momentum at $\lambda=0$ and the critical coupling $\lambda_c$, both scaled with the chemical potential $\mu$.}
\end{figure}

The formation of Fermi surfaces in holographic models was studied before in e.g. refs. \cite{Lee1,Zaanen1,Vegh1,Vegh2,JacobsUndoped}. Here we investigate how this formation depends on the parameters in our model, in particular on the size of the gap, i.e., on $\lambda$. In the spectral functions, Fermi surfaces appear as long-lived quasiparticle states at the chemical potential, i.e., at $\omega=0$. In other words, they appear as poles in the low-temperature spectral function at $\omega=0$ and at a nonzero Fermi momentum $k=k_F$. Examples of such spectral functions containing a Fermi surface are shown in figure \ref{fig:muspechol} in both standard and alternative quantization. In this figure we have set $\mu=2$ and $\lambda=1$. The Fermi surface is most clearly visible in alternative quantization. For a Fermi surface to appear in our model we need a sufficiently large chemical potential. This is of course a consequence of the gaps in our spectra. Moreover, we expect that since for larger values of $\lambda$ the gap grows, the chemical potential required for the formation of a Fermi surface will be higher. Conversely, for a fixed chemical potential there exists a critical coupling $\lambda_c$ at which the Fermi surface vanishes. This can indeed be seen in figure \ref{fig:kFhol}, where we plot the Fermi momentum of the spectral function in alternative quantization as a function of the coupling $\lambda$. From figure \ref{fig:kFholsc} it follows that when scaling the Fermi momentum with its value at $\lambda=0$, denoted by $k_{F,0}$, the resulting curve is given by
\be \label{eq:kFscale}
\frac{k_F}{k_{F,0}}=\sqrt{1-\frac{\lambda}{\lambda_c}}.
\ee
 \\
Hence, this curve is independent of $\mu$. Interestingly, this result is very reminiscent of a second-order mean-field quantum phase transition between a state with a Fermi surface and a state without one, if we think of the Fermi momentum as an order parameter. A quantum phase transition between such states was also found in the semiholographic model studied in ref. \cite{JacobsUndoped} and in ref. \cite{Liu1} in the context of nodal-line semimetals. Furthermore, the result looks similar to the result in non-interacting Dirac theory where $k_F=\sqrt{\mu^2-m^2}$, with $m$ the Dirac mass of that theory. This suggests that also here we can approximate the band in figure \ref{fig:SpFmuhol} by $\omega+\mu=\sqrt{k^2+m^2}$, up to some mass and wavefunction renormalization. Identifying the mass in our model with $\lambda$,\footnote{This identification can be justified by figure \ref{fig:wpeak}.} this seems to suggest that $k_{F,0}\propto\mu$ and $\lambda_c\propto\mu$. As shown in figure \ref{fig:kF0gc}, this is indeed what we obtain from our numerics when $\mu$ is large enough. For smaller chemical potential the mass and wavefunction renormalization still depend on $\mu$. Notice that the intersection of the curves in \ref{fig:kF0gc} with the vertical axis occurs at nonzero values, which indicates that also for small chemical potentials $k_{F,0}$ and $\lambda_c$ are proportional to $\mu$ to leading order.

Another feature that is visible in the spectral function in figure \ref{fig:SpFmuhol} is a second band that is close to the chemical potential, i.e., near $\omega=0$. Here the height of the peak is higher for lower values of $k$. A similar feature was observed in the spectral functions in ref. \cite{Lee1}, where the band was interpreted as a so-called critical Fermi ball. Moreover, refs. \cite{Vegh1,Vegh2} report the formation of multiple Fermi surfaces in their models. However, in our model, the imaginary part of the Green's function near the second band is nonzero, so this does not cause additional Fermi surfaces. Since the band is situated at the chemical potential, independent of the value of $\lambda$, it appears to indicate the existence of a many-body fermionic bound state in the theory due to the presence of the Fermi surface that hybridizes with the composite fermions described by the spectral function. We have found that the band is also present for nonzero values of $\mu$ that are small enough such that the band is inside the gap. Therefore the binding energy of this bound state, as compared to the band gap, can be positive or negative in our models. 
\begin{figure}[!t]
  \begin{subfigure}[b]{.44\linewidth}
	\centering 
	\includegraphics[clip=true, scale=.42]{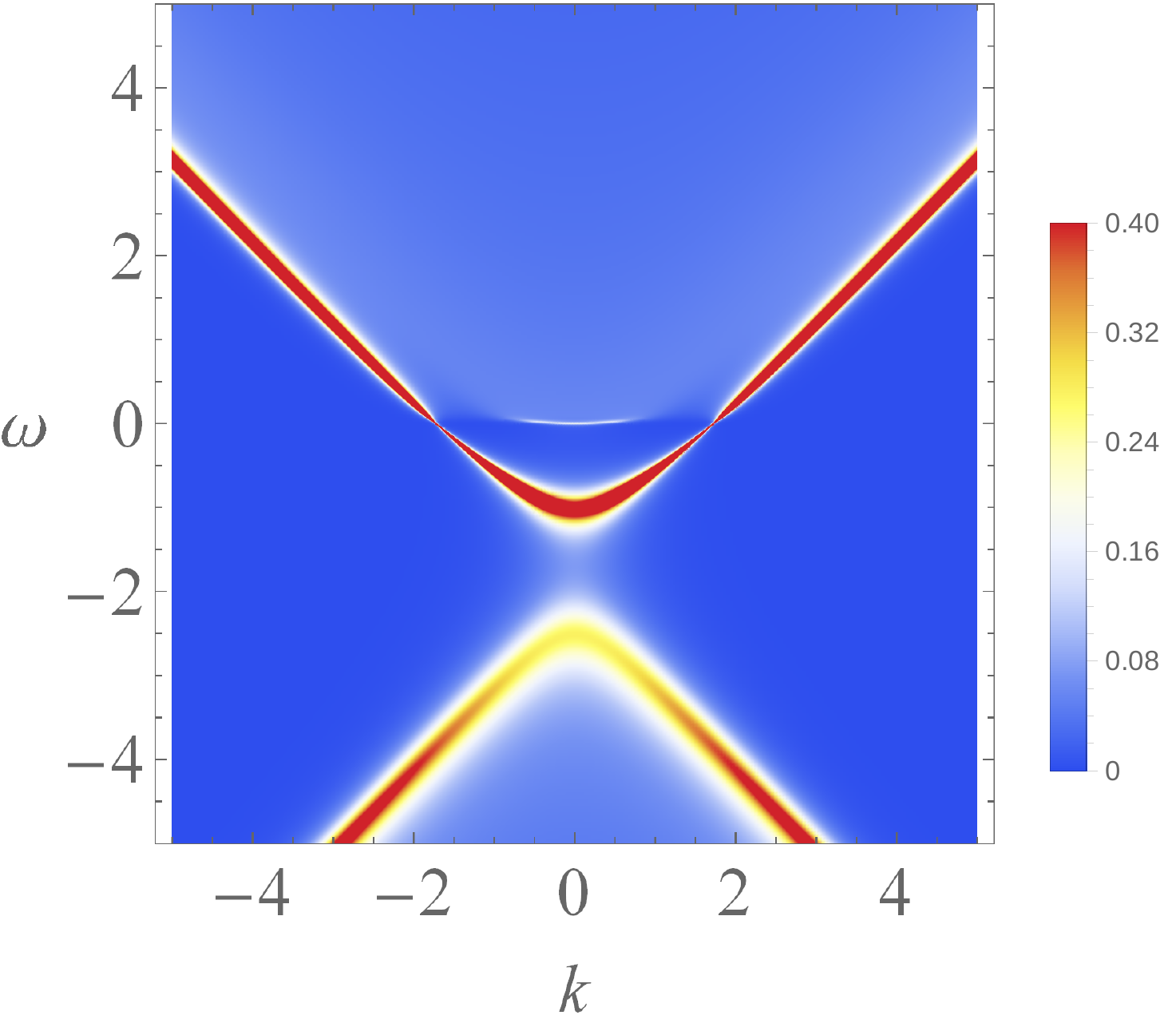}
		    \caption{}
		\label{fig:SpFmu}
  \end{subfigure}
  \begin{subfigure}[b]{.56\linewidth}
	\includegraphics[clip=true, scale=.434]{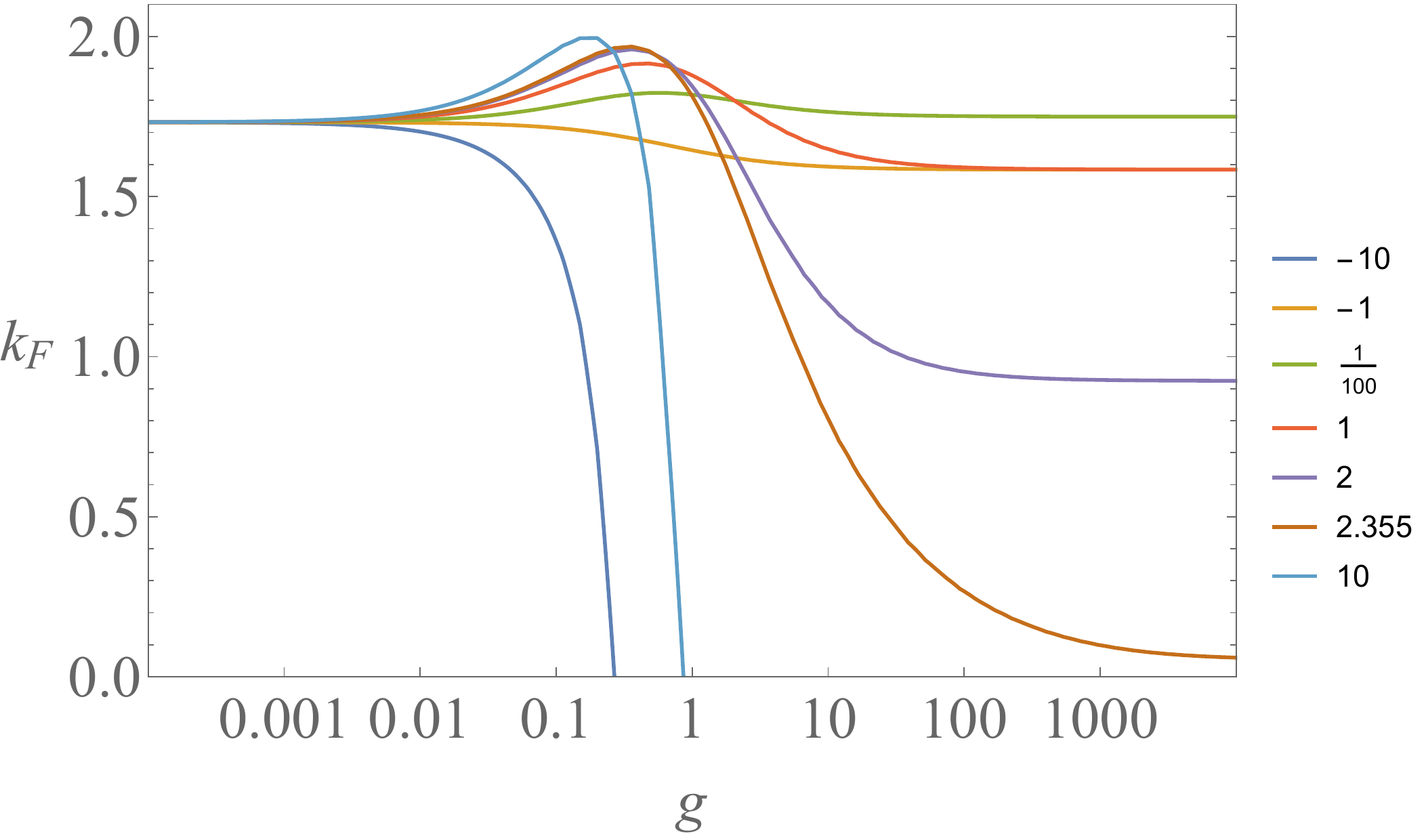}	
		 \caption{}
		\label{fig:kF}
  \end{subfigure}
\caption{\label{fig:muspec} (a) The spectral function for $\mu=2$, $\lambda=1$ and $g=4$. There is a Fermi surface at $k_F=1.72$. (b) The Fermi momentum $k_F$ as a function of $g$. The legend shows the value of the coupling $\lambda$.}
\end{figure}
\subsubsection{Semiholographic spectra} \label{sss:semispec}
Fermi surfaces in semiholographic models have been studied before in refs. \cite{Faulkner1,JacobsUndoped,Policastro2}. Here, we investigate the influence of a finite semiholographic coupling $g$ on our fermionic spectra. The semiholographic spectral function is shown in figure \ref{fig:SpFmu} for $g=4$ and otherwise the same parameters as in figure \ref{fig:muspechol}. Qualitatively the spectrum bears resemblance to the holographic spectrum in alternative quantization in figure \ref{fig:SpFmuhol}, but we observe quantitative differences in the size of the gap and the location of the Fermi surface. Figure \ref{fig:kF} shows the dependence of the Fermi momentum on both the couplings $\lambda$ and $g$. Here we have taken $\mu=2$, such that at $g=0$ there is a Fermi surface located at $k_F=\sqrt{\mu^2-M_0^2}=\sqrt{3}$, independent of $\lambda$. Furthermore, in the limit of large $g$ the Fermi momentum converges to the holographic value shown in figure \ref{fig:kFhol}. Since this value is independent of the sign of $\lambda$,  the Fermi momenta for $\lambda=\pm 1$ converge to the same value. Moreover, as we have seen in the holographic case, the Fermi surface disappears for large values of $|\lambda|$, but as seen in the figure, the value at which it disappears increases as we move away from the holographic limit. Finally, we note that for $\lambda>0$ the Fermi momentum first increases as we increase $g$, after which it decreases to ultimately converge to the large-$g$ value. Since the Fermi momentum is largest when the effective mass vanishes, the tops of these curves correspond to the quantum critical points $g_c$ mentioned in section \ref{ss:undoped}, where the spectrum is gapless. The value of $g_c$ can however not be derived from an equation similar to \eqref{eq:gcrit}, since $\Xi_c(k_\mu=0)$ is not purely imaginary for nonzero chemical potential. In contrast, for $\lambda<0$ the Fermi momentum converges to the large-$g$ value without passing through such a quantum critical point, in accordance with the findings in section \ref{ss:undoped}. 

Above we considered a case where $\mu>1$. Here, we have seen that for $g=0$ there is a Fermi surface. In contrast, in the limit $g\rightarrow\infty$, this is only true if $|\lambda|<\lambda_c$, where $\lambda_c$ can be determined from the asymptotic value in figure \ref{fig:kF0gc}. For finite $g$, this generalizes to two $g$-dependent critical values $\lambda_{c,+}(g)>0$ and $\lambda_{c,-}(g)<0$, so that the spectrum only contains a Fermi surface when $\lambda_{c,-}<\lambda<\lambda_{c,+}$. From figure \ref{fig:kF} we see that $\lambda_{c,+}$ increases as we lower $g$, and diverges to $\infty$ as $g\rightarrow 0$. Moreover, as we can see by comparing the curves for $\lambda=\pm 10$, we in general have that $\lambda_{c,-}\neq -\lambda_{c,+}$. The case when $\mu\leq 1$ is slightly more complicated. Here, there is no Fermi surface for the noninteracting case $g=0$. However, we know from figure \ref{fig:kF0gc} that there is a Fermi surface when $g\rightarrow\infty$ for $|\lambda|$ sufficiently small, i.e., $|\lambda|<\lambda_c$. This means that in this case there must be a critical coupling $g$ above which a Fermi surface forms. On the other hand, for larger $|\lambda|$, there is no Fermi surface when $g=0$ and neither when $g\rightarrow \infty$. In this case, depending on the value of $\lambda$, a Fermi surface can form at an intermediate value of $g$, which disappears again for larger $g$. Alternatively, there can also be values of $\lambda$ for which there is no Fermi surface for any value of $g$.

Besides the critical points mentioned above, we can find critical couplings $g_c$ reminiscent of the ones mentioned section \ref{ss:undoped} at which the spectrum is gapless. However, due to the nonzero imaginary part of the self-energy at nonzero doping, the spectrum will in this case be gapless for a larger range of values for $g$. Although interesting, showing the entire phase diagram with the all of the abovementioned critical couplings for all values of $g$, $\lambda$ and $\mu$ is beyond the scope of the present paper. 

The semiholographic spectral functions computed in this section obey the sum rule. This allows us to define and compute the momentum distribution function, which we shall do next.
\subsubsection{Momentum distribution functions}
\begin{figure}[!t]
  \begin{subfigure}[b]{.5\linewidth}
	\centering 
	\includegraphics[clip=true, scale=.47]{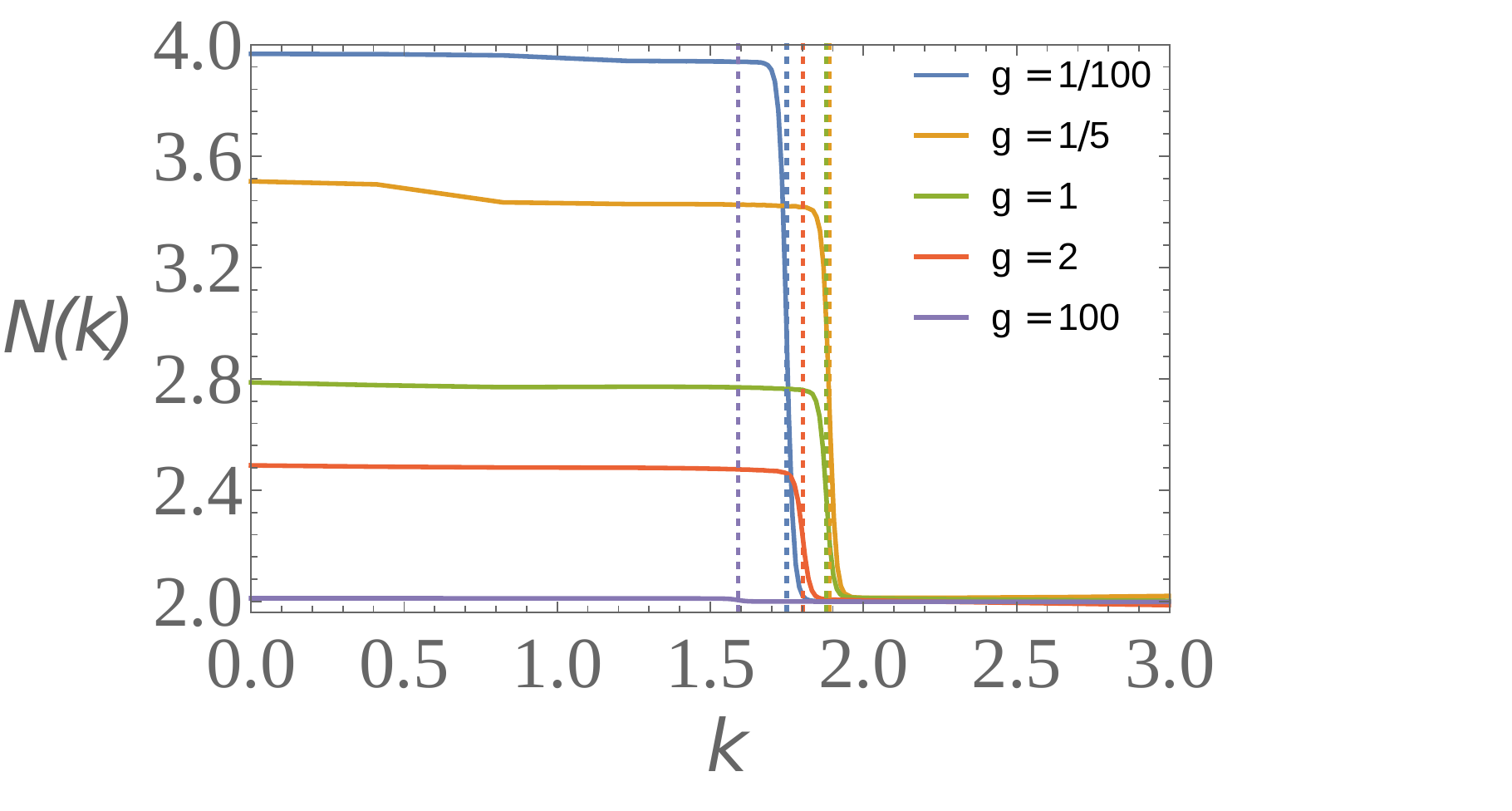}
		    \caption{}
		\label{fig:MDg}
  \end{subfigure}
  \begin{subfigure}[b]{.5\linewidth}
	\includegraphics[clip=true, scale=.474]{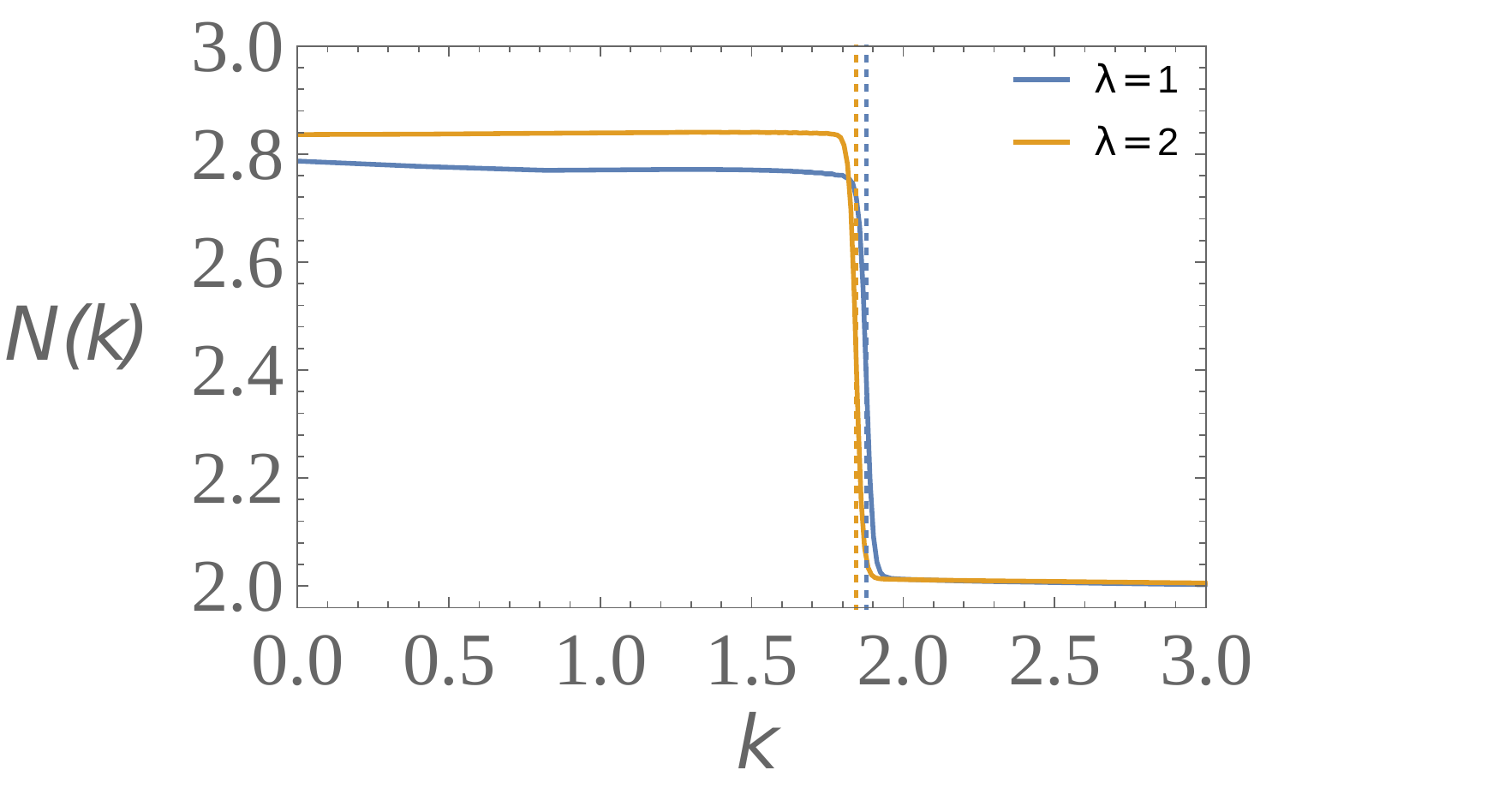}	
		 \caption{}
		\label{fig:MDgc}
  \end{subfigure}
\\ \\
  \begin{subfigure}[b]{.5\linewidth}
	\centering 
	\includegraphics[clip=true, scale=.515,trim=0cm 0cm 3cm 0cm]{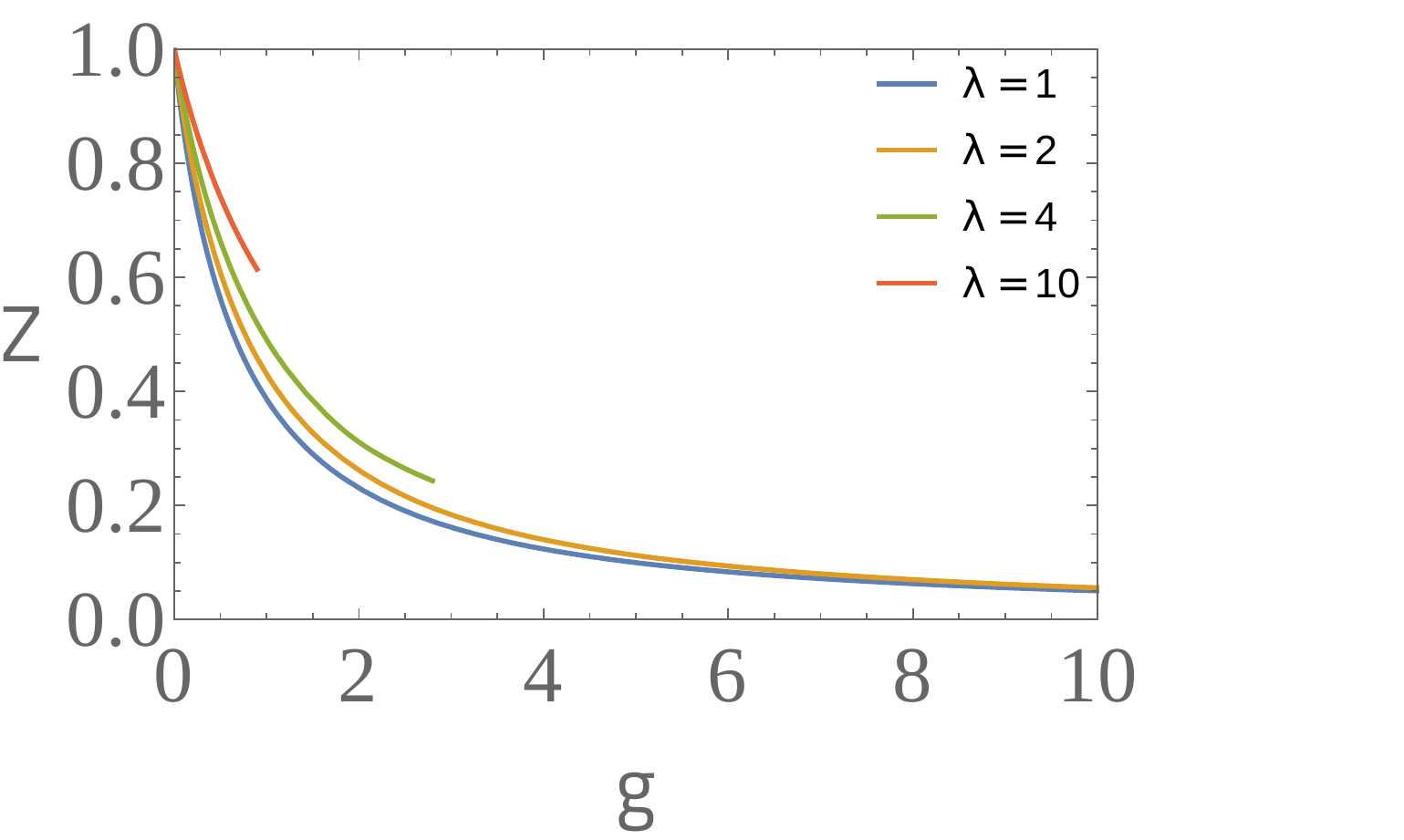}
		    \caption{}
		\label{fig:Z}
  \end{subfigure}
  \begin{subfigure}[b]{.5\linewidth}
	\includegraphics[clip=true, scale=.48,trim=0cm 0cm 0cm 0.2cm]{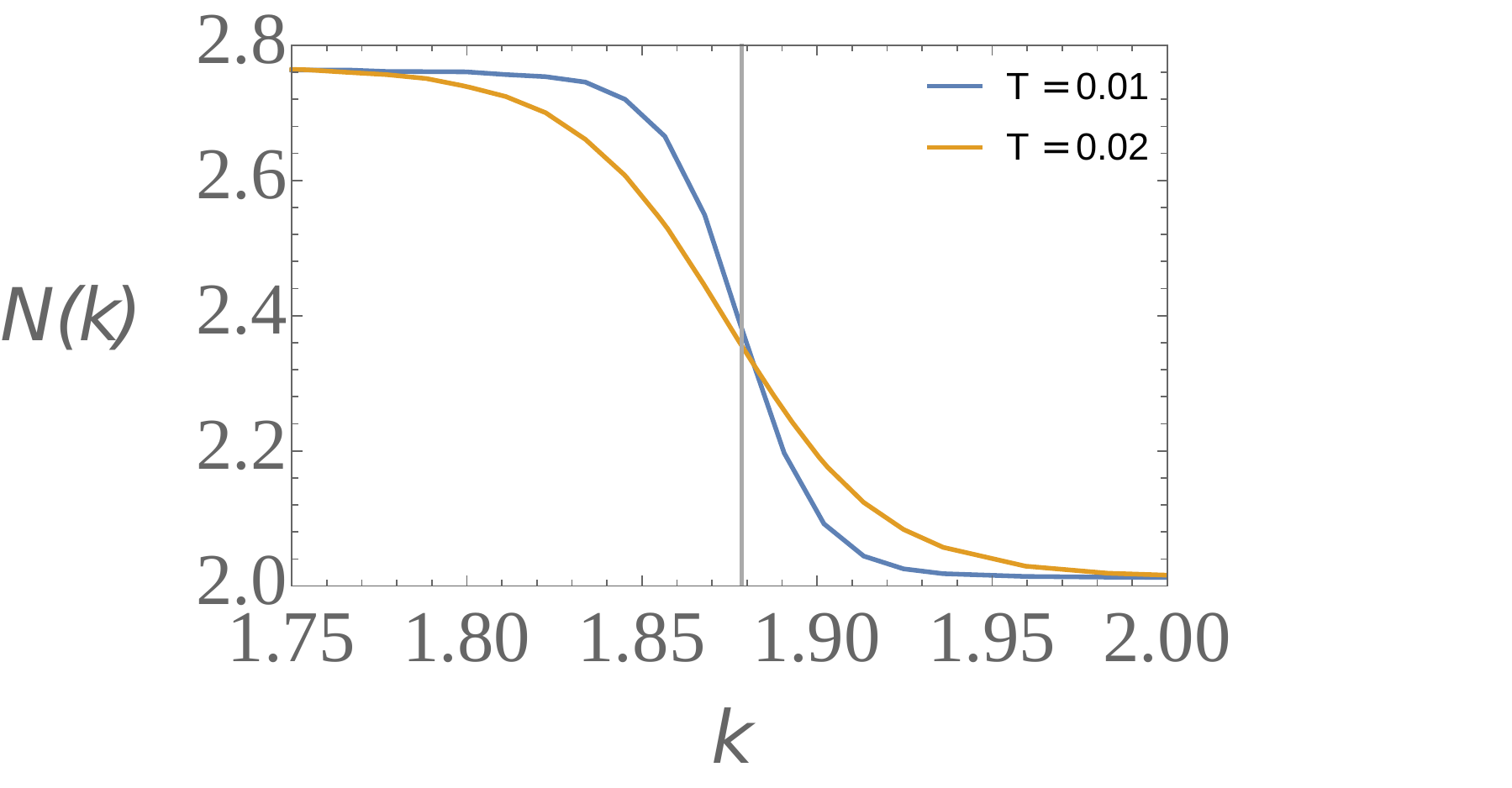}	
		 \caption{}
		\label{fig:MDT}
  \end{subfigure}
\caption{\label{fig:MDs} The dependence of the momentum distribution function on (a) the coupling $g$, (b) the coupling $\lambda$ and (d) the temperature, and (c) the dependence of the quasiparticle residue on $g$ and $\lambda$. In all plots the chemical potential is fixed to $\mu = 2$ and in all plots except for (d), the temperature is fixed to $T=1/100$. In (a) $\lambda = 1$, in (b) $g=1$ and in (d) $g=\lambda=1$. The dotted lines in (a) and (b) denote the Fermi momenta of the corresponding curves of the same colors, as calculated from the spectral functions. In (d), the gray line denotes the location of the Fermi surface, which does not depend on temperature.}
\end{figure}
Since our numerical calculations are performed at a small but nonzero temperature, the peaks at the Fermi surface have a small finite width. Therefore the previously obtained values of $k_F$, which is a quantity defined at zero temperature, are formally only approximately at the Fermi surface. Another indication of the Fermi surface is found by looking for a discontinuity in the momentum distribution of the Dirac fermion, which is defined as
\be \label{eq:MD}
N(\vec{k})=\int\dd\omega \rho(\omega,\vec{k})n_F(\omega)
\ee
where $n_F(\omega)$ is the Fermi distribution function. In figures \ref{fig:MDg}, \ref{fig:MDgc} and \ref{fig:MDT} we have studied the dependence of this quantity on the parameters of our system. Here we have fixed the chemical potential to $\mu=2$, as we expect that this will not have a large impact on the qualitative behavior besides the location of the Fermi surface. As expected, these momentum distribution functions contain a discontinuity at the Fermi surface, although the discontinuity is smoothed out by the finite temperature. We have however checked that the discontinuity becomes steeper as we decrease the temperature further, as is shown in figure \ref{fig:MDT}. 

For small momenta, the value of $N(k)$ depends on the couplings $g$ and $\lambda$, as shown in figure \ref{fig:MDg} and \ref{fig:MDgc}. The large deviation from $4$ is indeed a signature of strong interactions. It indicates that there is still nontrivial spectral weight above the upper band which can for example be seen in figure \ref{fig:SpFmu}. For $g=100$ the spread of spectral weight is so large that the discontinuity is hardly visible in the figure. In contrast, for $g=1/100$, nearly all spectral weight is contained in the two peaks in the spectrum. Furthermore, for large momenta, $N(k)$ always approaches $2$, independent of $g$ and $\lambda$. This shows that even though the spectrum contains a lot of nonzero spectral weight besides the peaks, the weight is still evenly distributed between the region above and below the gap due to the particle-hole symmetry of the undoped system. The discontinuity in the distributions is related to the quasiparticle residue $Z$, which we study in more detail shortly. Since the momentum distribution as defined in \eqqref{eq:MD} is the momentum distribution function for the entire Dirac spinor, it contains both particle and hole degrees of freedom of both chiralities. In the free theory, a discontinuity of $2$ in the spectral weight corresponds to the spin degrees of freedom only, which corresponds to $Z=1$. More generally, the quasiparticle residue is equal to one half times the discontinuity in the momentum distribution due to spin degeneracy. Figure \ref{fig:Z} shows that $Z$ increases as $\lambda$ increases, and decreases for increasing $g$, in accordance with figures \ref{fig:MDgc} and \ref{fig:MDg}. Moreover, the curves corresponding to $\lambda=4$ and $\lambda=10$ in figure \ref{fig:Z} terminate at a finite value of $g$. This is because as explained in section \ref{sss:semispec}, when $\lambda$ is high enough there is a critical value of $g$ above which there is no Fermi surface.
\\ \\
The momentum distributions also slightly decrease at a momentum below $k_F$. This is for example clearly visible in figure \ref{fig:MDg} for the case $g=1/5$. The reason is that the spectral functions contain a second band above the gap, as was mentioned in the previous section and can also be seen in figure \ref{fig:SpFmu}. Since this band is situated around $\omega=0$, it yields a contribution to the momentum distribution for low momenta.
\\ \\
The quasiparticle residue displayed in figure \ref{fig:Z} is calculated using the self-energy near the Fermi surface. For this purpose, we linearize the theory around the Fermi momentum $k_F$ and $\omega=0$. Starting with \eqqref{eq:renGR} and defining the shorthand notation
\ba
\omega_r&\equiv \omega+\mu+g\Xi_0, \\
k_r&\equiv k_3-g\Xi_3, \\
M_r&\equiv M_0+ig\Xi_c,
\ea
we see that we can write the trace of the Green's function as
\be \label{eq:TrGR}
\frac{1}{2}\text{Tr}\, G_R=\frac{-1}{\omega_r-\sqrt{k_r^2+M_r^2}}-\frac{1}{\omega_r+\sqrt{k_r^2+M_r^2}}.
\ee
Notice that this differs from a trivial free fermionic Green's function, since in general $\omega_r$, $k_r$ and $M_r$ are all complex functions of $\omega$, $k_3$ and all other parameters in our model. On the other hand, we obtain an effective model by linearizing the theory near the Fermi surface, i.e.,
\be \label{eq:effGR}
\frac{1}{2}\text{Tr}\, G_R=\frac{-Z}{\omega-v_F(k_3-k_F)-i \Sigma_{\text{eff}}(\omega,k_3)}.
\ee
This defines the quasiparticle residue $Z$, as well as the Fermi velocity $v_F$, which are both real and positive. Moreover, it defines the effective self-energy $\Sigma_{\text{eff}}$ which is a real function of $\omega$ and $k_3$ that vanishes at the Fermi surface at zero frequency. We can compute expressions for these quantities by comparing \eqqref{eq:effGR} to \eqqref{eq:TrGR}. To do so we first note that the latter is dominated by the first term, since this contains a pole exactly at the Fermi surface.\footnote{For $\mu<0$ the second term would dominate.} Neglecting the second term then yields 
\ba
Z&\approx\frac{-1}{\left.\partial_\omega\text{Re}\left[\big(\frac{1}{2}\text{Tr\,}G_R(\omega,k_F)\big)^{-1}\right] \right|_{\omega=0}} \nonumber \\
&\approx\frac{1}{1+\left.\partial_\omega\text{Re}\left[g \Xi_0(\omega,k_F)-\sqrt{ \big(k_F-g\Xi_3(\omega,k_F)\big)^2+\big(M_0+ig\Xi_c(\omega,k_F)\big)^2}\right]\right|_{\omega=0} }.
\ea
This expression allows us to calculate the quasiparticle residue by calculating the self-energy near the Fermi surface and was used to create figure \ref{fig:Z}. Note that for $M_0=\Xi_c=0$ this reduces to the result in section 4.2.1. in ref. \cite{JacobsUndoped}. Moreover, note that the evaluation at $k_F$ gives an additional implicit dependence on $g$ and $\lambda$. However, since $k_F$ converges for large $g$, we can still see from this expression that then $Z\propto 1/g$, as we also observe in the figure. Deriving a similar expression for the Fermi velocity yields
\ba
v_F&\approx 2Z \left.\partial_{k_3}\text{Re}\left[\big(\text{Tr\,}G_R(0,k_3)\big)^{-1}\right] \right|_{k_3=k_F} \nonumber \\
&\approx-Z\left.\partial_{k_3}\text{Re}\left[g \Xi_0(0,k_3)-\sqrt{ \big(k_3-g\Xi_3(0,k_3)\big)^2+\big(M_0+ig\Xi_c(0,k_3)\big)^2}\right]\right|_{k_3=k_F}.
\ea
Finally, we have that the effective self-energy is given by
\ba
\Sigma_{\text{eff}}&\approx 2Z\text{Im}\left[\lr{\text{Tr\,}G_R}^{-1}\right] \nonumber \\
&\approx -Z\text{Im}\left[g \Xi_0-\sqrt{ \big(k_3-g\Xi_3\big)^2+\big(M_0+ig\Xi_c\big)^2}\right].
\ea
\begin{figure}[!t]
  \begin{subfigure}[b]{.32\linewidth}
	\centering 
	\includegraphics[clip=true, scale=.354]{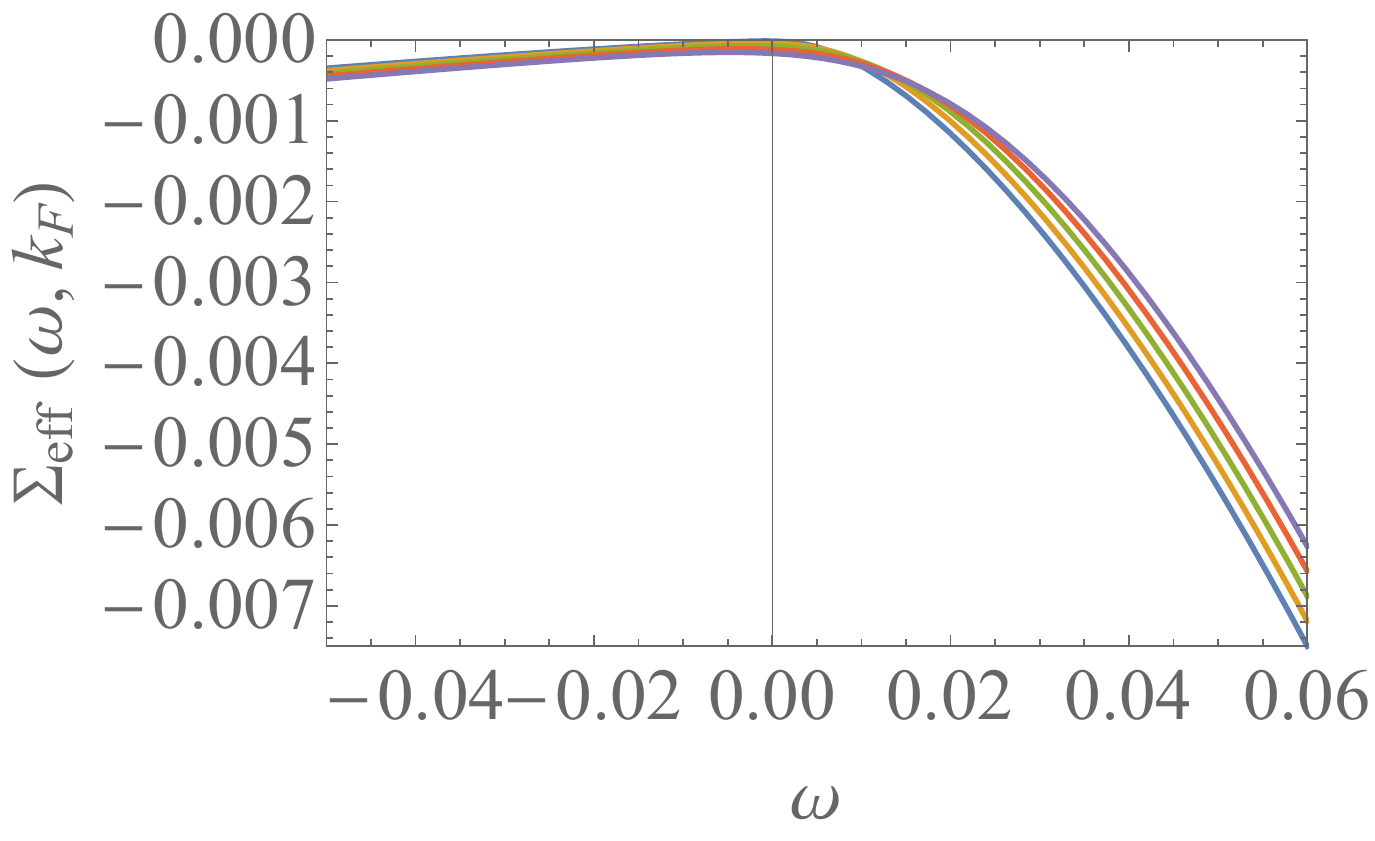}
		    \caption{}
		\label{fig:Sigmaeff}
  \end{subfigure}
  \begin{subfigure}[b]{.31\linewidth}
	\includegraphics[clip=true, scale=.344]{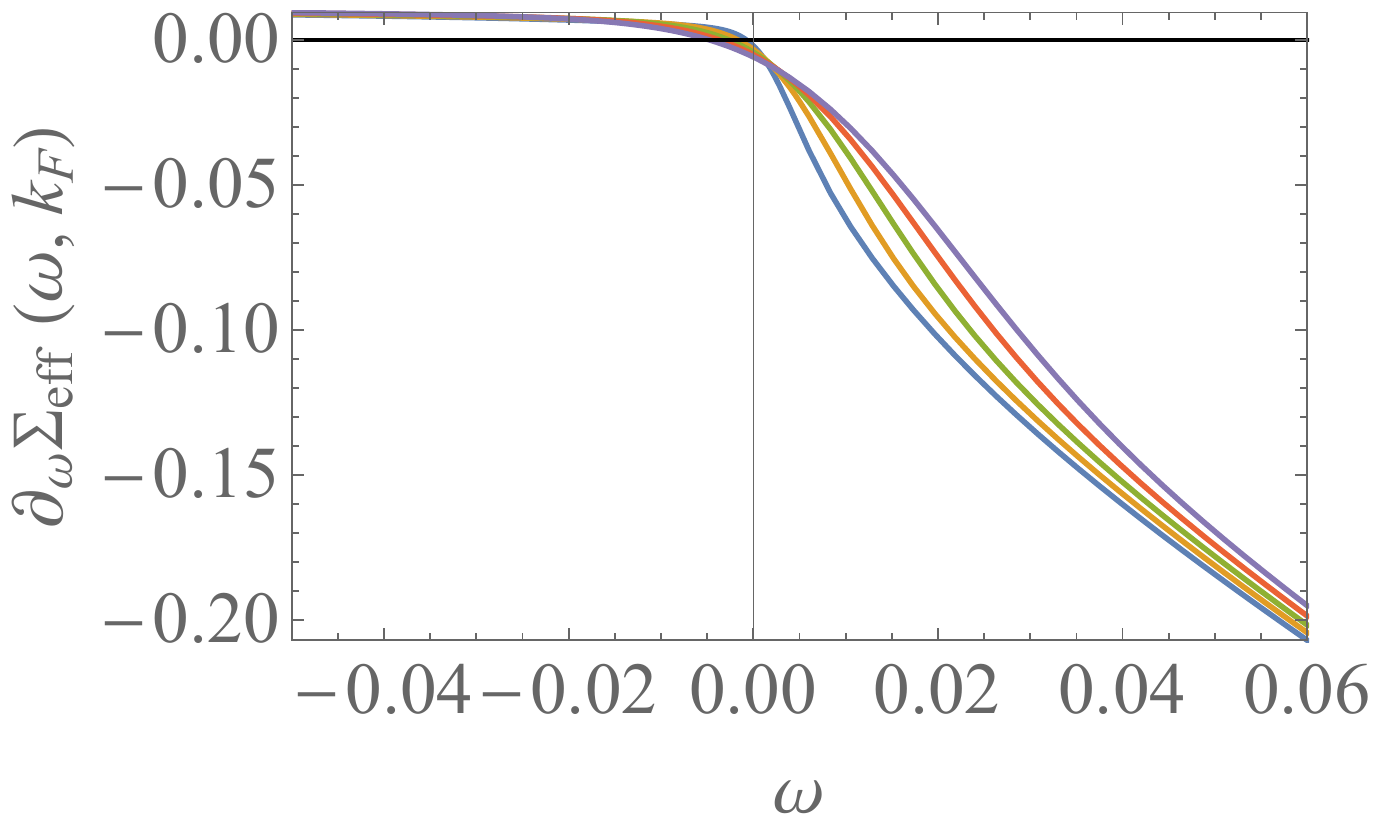}	
		 \caption{}
		\label{fig:dSigma}
  \end{subfigure}
  \begin{subfigure}[b]{.36\linewidth}
	\includegraphics[clip=true, scale=.35]{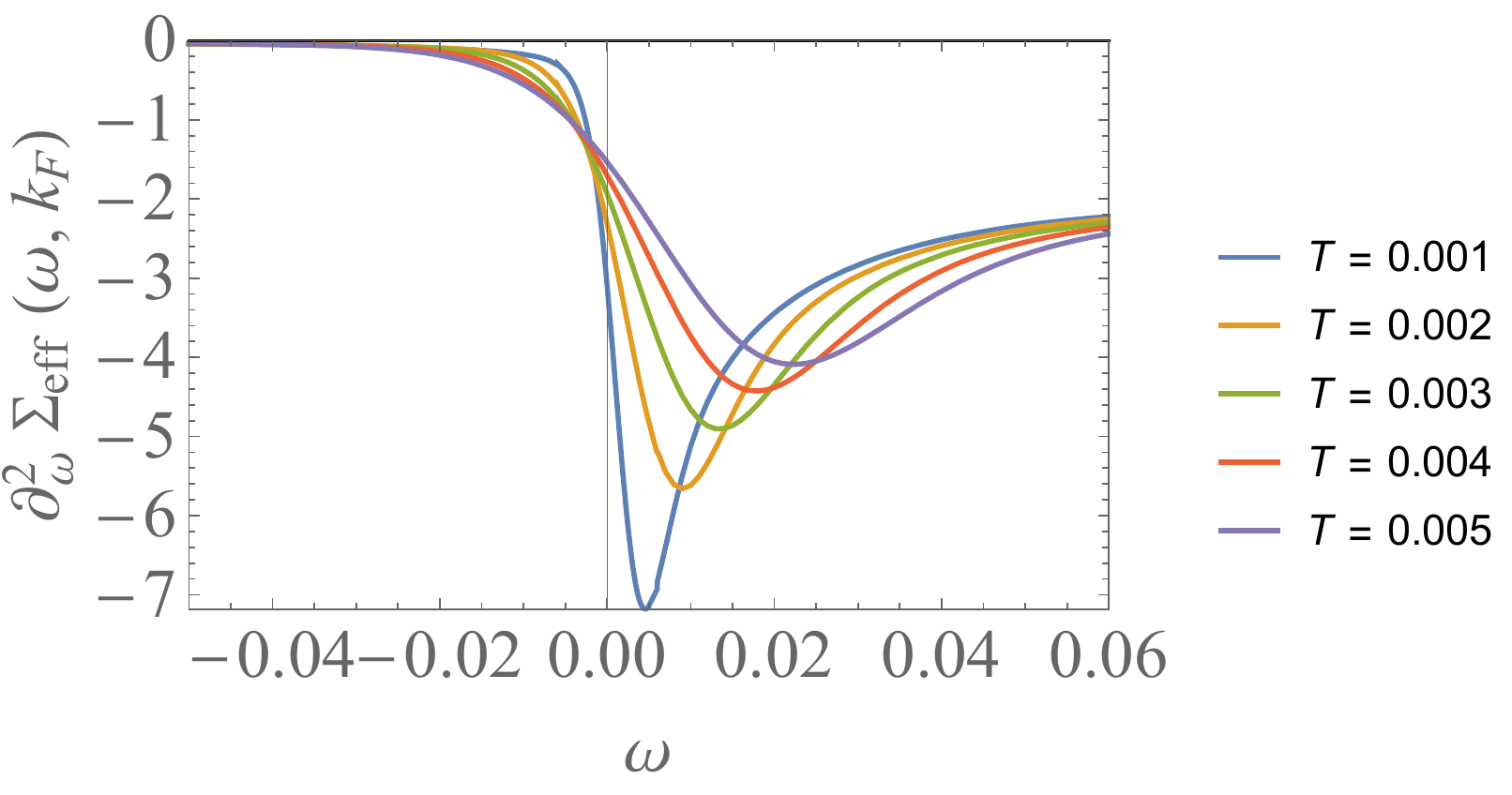}	
		 \caption{}
		\label{fig:ddSigma}
  \end{subfigure}
\caption{\label{fig:Sigmaeffds} (a) The effective self-energy $\Sigma_{\text{eff}}$ at the Fermi momentum, and its (b) first and (c) second derivatives. The legend on the right of (c) holds for all figures. In all graphs, $\mu=2$, $g=1$ and $\lambda=1$.}
\end{figure} \\
We have plotted this effective self-energy in figure \ref{fig:Sigmaeff}. As expected it vanishes at zero frequency,\footnote{In our numerics, it does not vanish exactly. We have however checked that this is due to nonzero temperature. As we lower the temperature, the dependence $\Sigma_{\text{eff}}(0,k_F)\propto T^{3/2}$ seems to approximate our results.} which is related to the fact at the Fermi surface there is a peak with an infinite lifetime $\tau = -2/\Sigma_{\text{eff}}$. This is indeed visible in figure \ref{fig:SpFmu}. At nonzero $\omega$, the width associated with the peak increases and so the lifetime becomes shorter. Of interest is the leading order frequency dependence of the self-energy $\Sigma_{\text{eff}}$. If the system behaves as a Fermi liquid, $\Sigma_{\text{eff}}$ should go to zero faster than $\omega$. Figure \ref{fig:dSigma} shows that this is indeed the case, since the derivative of $\Sigma_{\text{eff}}$ vanishes at $\omega=0$. Therefore the system indeed behaves like a Fermi liquid. In figure \ref{fig:ddSigma} we have also plotted the second derivative of $\Sigma_{\text{eff}}(\omega,k_F)$ for low temperatures. At first sight, these all seem to converge to a finite value as $\omega\rightarrow 0$, so that we could conclude that the self-energy follows $\Sigma_{\text{eff}}\propto \omega^2$ for small $\omega$. However, upon closer inspection, the coefficient of $\omega^2$ does not appear to converge as we lower the temperature. This convergence is necessary since we are studying the behavior near the Fermi surface, which is defined at zero temperature. The problem is that by determining the second derivative at zero frequency, we are inspecting a region where $\omega\lesssim T$, which does not correspond to the low-temperature physics. We do observe that the second derivative converges to a finite value for $\omega\approx 0.05$. Here, $T\ll \omega$, but $\omega$ is small compared to the other scales of the system, i.e., the gap and the chemical potential. We also see such a convergence for negative frequencies, but the convergence is to a different value. Although it is numerically difficult to lower the temperature even further, we believe that in the zero-temperature limit, the function in figure \ref{fig:ddSigma} converges to a finite value as $\omega\rightarrow 0$, which is dependent on whether the limit is taken from above or below. 

Although it appears difficult to extract the exact frequency dependence of the effective self-energy at the Fermi surface, it is at least clear that it converges to zero faster than linearly  and the system is a Fermi liquid. Moreover, since the second derivative clearly does not converge to 0, we can also conclude $\Sigma_{\text{eff}}(\omega,k_F)$ does not converge exponentially. We have checked that this does not change in the holographic limit of large $g$. This is in contrast to what is found in the models in \cite{JacobsUndoped,Hartnoll1,Iqbal1} where fermionic systems are studied using backgrounds with a dynamical scaling exponent $z$ that is emergent in the IR geometry. There, the self-energy behaves as
\be
\Sigma_{\text{eff}}\propto\exp\lr{-\lr{\frac{k_F^z}{\omega}}^{\frac{1}{z-1}}}.
\ee
Apparently, our model does not contain such an exponential behavior, as can also be checked by studying the IR geometry of our model in the zero-temperature limit, which is similar to the one in section 5.2 in ref. \cite{Horowitz0}.\footnote{Note however that this reference contains the geometry for $d=3$.}
\section{Conclusions and discussion}
In this work we provided a framework to study the dynamics of massive Dirac fermions using holographic and semiholographic models. We find that a gap is induced in the fermionic spectra by coupling two probe fermions in the bulk theory through a Yukawa coupling with the scalar field that provides the mass deformation. Moreover, by extending to a semiholographic model we encounter a quantum critical point at which the effective mass of the fermion vanishes. When turning on doping, we have seen that the Yukawa coupling can be used as a parameter that triggers a quantum phase transition between a state with a Fermi surface and a state without one. Studying the momentum distributions near the Fermi surfaces using the semiholographic Green's functions revealed that the described systems show Fermi-liquid behavior. In particular, the effective self-energy at the Fermi momentum converges to zero faster than linearly in frequency, as is expected from Pauli blocking.

An aspect that requires further research is the conductivity in the model. In this work, we have only seen that the CFT conductivity does not behave like an insulator despite the introduced mass deformation. It would be interesting to study if the fermionic contribution to the conductivity does behave like an insulator as expected. To this end we should first use the dressed semiholographic fermionic Green's functions to calculate the current-current correlation function from the one-loop diagram as in refs. \cite{Jacobs1,Jacobs3}. The next step would then be to also include the vertex corrections.

The principal result of this work is to provide a framework that allows us to compute fermionic spectral functions that are relevant in the study of strongly coupled condensed matter. In particular, using the method described here allows us to include the mass of the fermions in condensed-matter systems. As mentioned in the introduction, this can either be a real particle mass or an effective mass or gap in a Dirac material. The viewpoint taken in this work is the latter, where the Dirac theory is used as an effective description. Basically, this means that the speed $c$ that is set to unity in this paper corresponds to the Fermi velocity of the free massless Dirac theory. From the other point of view, we use Dirac theory to describe actual fermionic particles, and the speed $c$ corresponds to the speed of light. In this case the newly introduced energy scale $mc^2$, where $m$ is the mass of the fermion in the condensed-matter system under study, allows us to research nonrelativistic physics by resorting to the appropriate regime in which the other scales such as $k_BT$, $\hbar\omega$ and $\mu$ are small compared to $mc^2$. It would be interesting to see if we can use this approach to for example study ultracold Fermi gases at unitarity. Besides this, we can of course study a plethora of strongly coupled condensed-matter physics by adding additional ingredients in the bulk theory, such as chiral-symmetry breaking terms as was done in refs. \cite{Landsteiner1,Landsteiner2,Liu1}.

\acknowledgments

It is a pleasure to thank U. G\"{u}rsoy, S. Vandoren and E. Mauri for very helpful discussions, and the last two also for providing comments on this manuscript. This work was supported by the Stichting voor Fundamenteel Onderzoek der Materie (FOM) and is part of the D-ITP consortium, a program of the Netherlands Organisation for Scientific Research (NWO) that is funded by the Dutch Ministry of Education, Culture and Science (OCW).

\appendix
\section{Conventions} \label{app:conv}
In this appendix we list our conventions. We firstly specify the dimensionless units used throughout the paper, and subsequently give our conventions regarding the Dirac theory.
\subsection{Units} \label{app:units}
The action for the gravitational background \eqqref{eq:ActionBR} reads in SI units
\be \label{eq:ActionSI}
S=\int\dd^{5}x\sqrt{-g}\lr{\frac{c^3}{16\pi G}(R-2\Lambda)-\frac{1}{4{\mu_0} c}F^2- \lr{\lr{\partial\phi}^2+\frac{m_\phi^2c^2}{\hbar^2}\phi^2}}\,.
\ee
Here $G$ and $\mu_0$ are Newton's constant and the vacuum permeability respectively in 4 spatial dimensions, and $\Lambda<0$ is the cosmological constant. The components of the metric $g_{\mu\nu}$ are dimensionless. Defining the anti-de Sitter radius as $L^2=6/(-\Lambda)$, the dimensionless units in this paper are obtained by scaling length scales by $L$, so that  $\Lambda={-6}$. Moreover, we put Boltzmann's constant $k_B=1$. Consequently, all energy scales, such as $k_B T$, $m_\phi c^2$ and $M_0c^2$, are in units of $\hbar c/L$. The dimensionless fields are obtained as follows:
\ba \label{eq:dimlessfields}
\tilde{A}_{\tilde{t}}&=\sqrt{\frac{16\pi G}{\mu_0 c^6}}A_t, \\
\tilde{\phi}&=\sqrt{\frac{16\pi G}{c^3}}\phi.
\ea
Here the left-hand sides are the dimensionless fields used throughout the paper, where we omitted the tildes. 

The Dirac action in \eqqref{eq:actionchiral} in SI units reads
\be 
S=ig_f \intbulk \bar{\psi}\lr{\slashed{D}-\frac{Mc}{\hbar}}\psi+ig_f\intbdy \bar{\psi}_R\psi_L.
\ee
Taking $g_f$ dimensionless, the Dirac field $\psi$ has the dimension of $\sqrt{\hbar}/L^2$. The dimensionless Dirac field can be defined by extracting this factor from the field. Alternatively, there is always an undetermined dimensionless constant $\hbar c^3/16\pi GL^3$ which can be included in the definition. However, this is equivalent to a redefinition of $g_f$. Furthermore, the dimensionless charge $\tilde{q}$ that resides in the covariant derivative is given by
\be
\tilde{q}=\sqrt{\frac{\mu_0 c^6}{16\pi G}}\frac{L}{\hbar c}q.
\ee
We remind the reader that the dimensionless quantities defined in this appendix are dimensionless in SI units, but can still have a nonzero scaling dimension.
\subsection{Dirac theory} \label{app:Dirac}
Firstly, we define Dirac's gamma matrices in flat spacetime as follows:

\begin{align}
		\Gamma^{\underline{\mu}} &= \gamma^{\underline{\mu}} = \begin{pmatrix} 0 & \bar{\sigma}^\mu \\ \sigma^\mu & 0 \end{pmatrix} \qquad \text{for $\mu\neq r$}, \\
		\Gamma^{\underline{r}} &= \gamma^{\underline{5}} = \begin{pmatrix} \mathbb{I}_2 & 0 \\ 0 & -\mathbb{I}_2 \end{pmatrix} = i \gamma^{\underline{0}} \gamma^{\underline{1}} \gamma^{\underline{2}} \gamma^{\underline{3}}.
\end{align}
Here $\Gamma^{\underline{a}}$ are the gamma matrices defined in the $4+1$ spacetime of the bulk, while $\gamma^{\underline{a}}$ are the usual gamma matrices in $3+1$ dimensions. Notice as well that these indices, such as $\underline{a}$, are underlined, meaning that these are tensors defined in flat spacetime, i.e., $g^{\underline{ab}} = \eta^{\underline{ab}}$. Moreover, $\sigma=(\mathbb{I}_2,\sigma^i)$ and $\bar{\sigma}=(-\mathbb{I}_2,\sigma^i)$ with $\sigma^i$ the Pauli matrices.

Conjugate spinors are defined as
\be
\bar{\psi}=\psi^\dagger \Gamma^{\underline{0}}.
\ee
The gamma matrices in a curved background are defined using the vielbeins:
\begin{equation}
	\Gamma^\mu = \vielbein{\mu}{a} \Gamma^{\underline{a}}.
\end{equation}
The vielbeins satisfy $g_{\mu\nu}=e^{\underline{a}}_\mu e^{\underline{b}}_\nu \eta_{{\underline{a}}{\underline{b}}}$ and the inverse vielbeins $e^\mu_{\underline{a}}e^{\underline{a}}_\nu=\delta_\nu^\mu$ and $e^\mu_{\underline{a}}e^{\underline{b}}_\mu=\delta_{\underline{a}}^{\underline{b}}$. Computing these for the metric in \eqqref{eq:metricAnsatz} gives

\begin{align}
		\vielbein{0}{0} &= \sqrt{\frac{e^{\chi(r)}}{f(r)}}, \\
		\vielbein{r}{r} &= \sqrt{f(r)}, \\
		\vielbein{i}{i} &= \frac{1}{r}.
\end{align}
The slash is defined by 
\be
\slashed{X}=\Gamma^\mu X_\mu.
\ee
Finally, the Dirac action contains the covariant derivative $\slashed{D}=\slashed{\nabla}-iq\slashed{A}$. Here the spinor covariant derivative $\nabla_\mu$ is defined as
\be
\nabla_\mu\psi=\partial_\mu\psi+\Omega_\mu \psi
\ee
where $\Omega_\mu$ is given by
\be
\Omega_\mu=\frac{1}{8}\omega_{\mu\underline{ab}}[\Gamma^{\underline{a}},\Gamma^{\underline{b}}]
\ee
with $\omega_{\mu\underline{ab}}$ the spin connection
\be
\omega^{\underline{a}}_{\mu\underline{b}}=e^{\underline{a}}_\nu e^{\lambda}_{\underline{b}}\Gamma^\nu_{\mu\lambda}-e^\lambda_{\underline{b}}\partial_\mu e^{\underline{a}}_\lambda.
\ee
For the metric in \eqqref{eq:metricAnsatz} the only nonvanishing components of the spin connection are $\omega_{t\underline{tr}}=-\omega_{t\underline{rt}}$ and $\omega_{i\underline{ir}}=-\omega_{i\underline{ri}}$. Using this, one can show that the spinor covariant derivative can be written as
\be
\slashed{\nabla}\psi=\slashed{\partial}\psi+\Gamma^{\underline{r}} F(r) \psi
\ee
where $F$ is a function depending on the radial coordinate only. Defining
\be
p(r)=\exp\lr{-\int^r \dd r' F(r')},
\ee
we have that
\be \label{eq:rescale}
\slashed{\nabla}\lr{p\psi}=p\slashed{\partial}\psi.
\ee
This demonstrates that rescaling both $\psi^{(1)}\rightarrow p \psi^{(1)}$ and $\psi^{(2)}\rightarrow p \psi^{(2)}$ in eqs. \eqref{eq:Dirac1} and \eqref{eq:Dirac2} gets rid of the spin connection terms in the Dirac equations. Moreover, this rescaling does not affect the matrix $\Xi$ defined in \eqqref{eq:defXi}.
\section{Symmetries} \label{app:sym}
In certain cases we can reduce the amount of equations we need to solve by using additional symmetries. From the equations in \eqref{eq:redXieqs} and the imposed initial conditions we can derive that
\ba
\Xi_\pm(\tilde{\omega},k_3)&=\Xi_\mp(\tilde{\omega},-k_3), \label{eq:symkpm}\\
\Xi_c(\tilde{\omega},k_3)&=\Xi_c(\tilde{\omega},-k_3), \label{eq:symkc} \\
\Xi_\pm(\tilde{\omega},k_3)&=-\Xi^*_\mp(-\tilde{\omega},k_3), \label{eq:symomegapm} \\
\Xi_c(\tilde{\omega},k_3)&=-\Xi^*_c(-\tilde{\omega},k_3). \label{eq:symomegac}
\ea
The first two symmetries correspond to parity symmetry, whereas the last two represent time-reversal symmetry. From these symmetries it follows that we can solve the system for $k_3\geq 0$ and use \eqqref{eq:symkpm} and \eqqref{eq:symkc} to obtain the results for $k_3<0$. Moreover, when $\mu=0$, we only need to solve for $\omega\geq 0$ according to \eqqref{eq:symomegapm} and \eqqref{eq:symomegac}, which then represent particle-hole symmetry.

Also, the equations in \eqref{eq:redXieqs} as well as the initial conditions are invariant under sending both $\lambda\rightarrow -\lambda$ and $\Xi_c\rightarrow -\Xi_c$. This symmetry allows us to obtain solutions for $\lambda<0$ from solutions with $\lambda>0$. Furthermore, multiplying the matrix equation between the brackets in \eqref{eq:Xieq} from both the left and the right by $\Xi^{-1}$ reveals that the equations are also invariant under sending $M \rightarrow -M$, $\lambda \rightarrow -\lambda$ and $\Xi \rightarrow \Xi^{-1}$. Consequently, solutions for $M<0$ can be obtained from solutions with $M>0$. Finally, solutions for $\mu<0$ can be obtained from solutions with $\mu>0$ by exploiting the symmetries in \eqqref{eq:symomegapm} and \eqqref{eq:symomegac}.

The symmetries in eqs. \eqref{eq:symkpm}-\eqref{eq:symomegac} show that for $k_3=0$ we have $\Xi_3=0$, and that for $\tilde{k}_\mu=0$, $\Xi_\pm$ is real and $\Xi_c$ is imaginary. This demonstrates that the constants $Z_0$ and $M_{\text{eff}}$ defined in \eqqref{eq:Z} and \eqqref{eq:Meff} respectively are real. 
\section{RG equations} \label{app:RG} 
Throughout the paper, we have fixed the ratio of the scalar source $\phi_s$ and the bare mass $M_0$ originating from the UV action \eqqref{eq:SUV} to $\alpha\equiv M_0/\phi_s=\sqrt[4]{\pi^2/3}$. Here we present an argument for choosing this specific value. It is important to realize that we could in principle consider models in which this ratio is a free parameter, so that independent from the argument made here, our semiholographic results can also be seen as a specific case of such models. Moreover, this value is not relevant for our holographic results.

For a mass $m_\phi^2=-3$, the asymptotic equations of motion in eqs. \eqref{eq:KG}, \eqref{eq:MW}, \eqref{eq:EE1} and \eqref{eq:EE2} give
\be \label{eq:exp1}
\phi=\phi_s r^{-1}+\phi_v r^{-3}-\frac{\phi_s^3}{6} r^{-3}\log r\dots\,.
\ee
Under a rescaling $r\rightarrow\lambda r$ we then get 
\be \label{eq:exp2}
\phi\rightarrow \phi_s(\lambda)\lambda^{-1}r^{-1}+\phi_v(\lambda)\lambda^{-3}r^{-3}-\frac{\phi_s(\lambda)^3}{6}\lambda^{-3}r^{-3}\lr{\log r+\log \lambda}\dots\,.
\ee
As $\phi$ is invariant under this rescaling we get from comparing \eqqref{eq:exp1} and \eqqref{eq:exp2} that
\ba
\phi_s(\lambda)&=\phi_s\lambda, \\
\phi_v(\lambda)&=\phi_v\lambda^3+\frac{\phi_s^3}{6}\lambda^3\log \lambda.
\ea
This yields the following RG equations for the coefficients $\phi_{s,v}$: 
\ba
\lambda\frac{\dd\phi_s(\lambda)}{\dd\lambda}&= \phi_s(\lambda), \label{eq:RGs1}\\
\lambda\frac{\dd\phi_v(\lambda)}{\dd\lambda}&=3\phi_v(\lambda)+\frac{\phi_s(\lambda)^3}{6}. \label{eq:RGs2}
\ea
Note that the term with the logarithm generates the nontrivial part of the RG equation in \eqqref{eq:RGs2}.

On the other hand, the two-point function found from semiholography is
\ba
\exv{i\bar{\Psi}\Psi}&=\text{Tr}\int \frac{\dd^4 k_E}{(2\pi)^4}\frac{1}{-i\slashed{k}_E+M_0-i\tilde{\Sigma}} \nonumber \\
&=\text{Tr}\int \frac{\dd^4 k_E}{(2\pi)^4}\frac{1}{k_E^2+M_0^2+\tilde{\Sigma}^2-\{\slashed{k}_E,\tilde{\Sigma}\}}\lr{i\slashed{k}_E+M_0-i\tilde{\Sigma}},\label{eq:RGs3}
\ea
where we used the Euclidean momentum with $k_E^2=|\vec{k}|^2+k_4^2$ and $k_4=i\omega$ and where $\{.,.\}$ denotes the anticommutator. Note that the self-energy $\tilde{\Sigma}$ here differs from the self-energy $\Sigma$ defined in \eqqref{eq:Sigmadef} by a factor of $\Gamma^{\underline{0}}$, since we are calculating $\exv{\bar{\Psi}\Psi}$ rather than $\exv{\Psi^\dagger\Psi}$. However, for large momenta its components are also proportional to $k^{2M}$. 

Our goal is now to derive an RG equation for $\exv{i\bar{\Psi}\Psi}$. To this end, we integrate over a high-momentum shell for which $|k_E|\in\{\Lambda e^{-l},\Lambda\}$ where $\Lambda$ is a UV cut-off and $l>0$. We then look for the logarithmic UV divergence, which should yield the term in the RG equation that can be compared to the nontrivial term in \eqqref{eq:RGs2}. For the high momenta in the shell, the integrand in \eqqref{eq:RGs3} can be expanded as
\be
\exv{i\bar{\Psi}\Psi}_\Lambda=\dots+\text{Tr}\int_\Lambda \frac{\dd^4 k_E}{(2\pi)^4}\frac{1}{k_E^2}\lr{1-\frac{M_0^2+\tilde{\Sigma}^2-\{\slashed{k}_E,\tilde{\Sigma}\}}{k_E^2}}\lr{i\slashed{k}_E+M_0-i\tilde{\Sigma}},
\ee
where the dots denote the integration over lower momenta and the second integral is over the shell. Now it is clear that the logarithmic divergence resides in the term\footnote{Here we neglect the subtle case $M=0$.} proportional to $M_0^3/k_E^4$. To evaluate this term, we use $\dd^4 k_E=2\pi^2 k_E^3\dd k_E$ to get
\be
\exv{i\bar{\Psi}\Psi}_\Lambda=\dots-\frac{M_0^3}{2\pi^2 }\int_{\Lambda e^{-l}}^{\Lambda} \frac{\dd k_E}{k_E}=\dots-\frac{M_0^3l}{2\pi^2}\, ,
\ee
so that 
\be
\frac{\dd \exv{i\bar{\Psi}\Psi}}{\dd l}=\dots-\frac{M_0^3}{2\pi^2}.
\ee
Expressing this in terms of $\alpha$ as defined in the beginning of this appendix, and identifying $\phi_v=-\exv{i\bar{\Psi}\Psi}/\alpha$,\footnote{This follows from comparing the on-shell Klein-Gordon action to the mass deformation.} we obtain
\be
\frac{\dd \phi_v}{\dd l}=\dots+\frac{\alpha^4 \phi_s^3}{2\pi^2}\,.
\ee 
Comparing this with \eqqref{eq:RGs2} then yields $\alpha^4=\pi^2/3$.

% The bibliography will probably be heavily edited during typesetting.
% We'll parse it and, using the arxiv number or the journal data, will
% query inspire, trying to verify the data (this will probalby spot
% eventual typos) and retrive the document DOI and eventual errata.
% We however suggest to always provide author, title and journal data:
% in short all the informations that clearly identify a document.

\medskip
\bibliographystyle{JHEP}
\bibliography{Bibliography}

\providecommand{\href}[2]{#2}\begingroup\raggedright\begin{thebibliography}{10}

\bibitem{Maldacena1999}
J.~Maldacena, \emph{The large-{N} limit of superconformal field theories and
  supergravity},
  \href{https://doi.org/10.1023/A:1026654312961}{\emph{International Journal of
  Theoretical Physics} {\bfseries 38} (Apr, 1999) 1113--1133},
  [\href{https://arxiv.org/abs/hep-th/9711200}{{\ttfamily hep-th/9711200}}].

\bibitem{Witten1998}
E.~Witten, \emph{{Anti-de {S}itter space and holography}},
  \href{https://doi.org/10.4310/ATMP.1998.v2.n2.a2}{\emph{Adv. Theor. Math.
  Phys.} {\bfseries 2} (1998) 253--291},
  [\href{https://arxiv.org/abs/hep-th/9802150}{{\ttfamily hep-th/9802150}}].

\bibitem{Gubser1}
S.~Gubser, I.~Klebanov and A.~Polyakov, \emph{Gauge theory correlators from
  non-critical string theory},
  \href{https://doi.org/https://doi.org/10.1016/S0370-2693(98)00377-3}{\emph{Physics
  Letters B} {\bfseries 428} (1998) 105 -- 114},
  [\href{https://arxiv.org/abs/hep-th/9802109}{{\ttfamily hep-th/9802109}}].

\bibitem{Policastro1}
G.~Policastro, D.~T. Son and A.~O. Starinets, \emph{Shear viscosity of strongly
  coupled {$N\phantom{\rule{0ex}{0ex}}=\phantom{\rule{0ex}{0ex}}4$}
  supersymmetric {Yang-Mills} plasma},
  \href{https://doi.org/10.1103/PhysRevLett.87.081601}{\emph{Phys. Rev. Lett.}
  {\bfseries 87} (Aug, 2001) 081601},
  [\href{https://arxiv.org/abs/hep-th/0104066}{{\ttfamily hep-th/0104066}}].

\bibitem{Kovtun1}
P.~K. Kovtun, D.~T. Son and A.~O. Starinets, \emph{Viscosity in strongly
  interacting quantum field theories from black hole physics},
  \href{https://doi.org/10.1103/PhysRevLett.94.111601}{\emph{Phys. Rev. Lett.}
  {\bfseries 94} (Mar, 2005) 111601},
  [\href{https://arxiv.org/abs/hep-th/0405231}{{\ttfamily hep-th/0405231}}].

\bibitem{HHH1}
S.~A. Hartnoll, C.~P. Herzog and G.~T. Horowitz, \emph{Building a holographic
  superconductor},
  \href{https://doi.org/10.1103/PhysRevLett.101.031601}{\emph{Phys. Rev. Lett.}
  {\bfseries 101} (Jul, 2008) 031601},
  [\href{https://arxiv.org/abs/0903.3295}{{\ttfamily 0903.3295}}].

\bibitem{HartnollLec}
S.~A. Hartnoll, \emph{Lectures on holographic methods for condensed matter
  physics}, {\emph{Classical and Quantum Gravity} {\bfseries 26} (2009)
  224002}, [\href{https://arxiv.org/abs/0903.3246}{{\ttfamily 0903.3246}}].

\bibitem{McGreevy2009}
J.~McGreevy, \emph{Holographic duality with a view toward many-body physics},
  \href{https://arxiv.org/abs/0909.0518}{{\ttfamily 0909.0518}}.

\bibitem{Sachdev2011}
S.~Sachdev, \emph{Condensed Matter and {AdS/CFT}}, pp.~273--311.
\newblock Springer Berlin Heidelberg, Berlin, Heidelberg, 2011.
\newblock \href{https://arxiv.org/abs/1002.2947}{{\ttfamily 1002.2947}}.

\bibitem{Son1}
D.~T. Son, \emph{Toward an {AdS/cold} atoms correspondence: A geometric
  realization of the {Schr\"odinger} symmetry},
  \href{https://doi.org/10.1103/PhysRevD.78.046003}{\emph{Phys. Rev. D}
  {\bfseries 78} (Aug, 2008) 046003},
  [\href{https://arxiv.org/abs/0804.3972}{{\ttfamily 0804.3972}}].

\bibitem{McGreevy1}
K.~Balasubramanian and J.~McGreevy, \emph{Gravity duals for nonrelativistic
  conformal field theories},
  \href{https://doi.org/10.1103/PhysRevLett.101.061601}{\emph{Phys. Rev. Lett.}
  {\bfseries 101} (Aug, 2008) 061601},
  [\href{https://arxiv.org/abs/0804.4053}{{\ttfamily 0804.4053}}].

\bibitem{Kachru1}
S.~Kachru, X.~Liu and M.~Mulligan, \emph{Gravity duals of {L}ifshitz-like fixed
  points}, \href{https://doi.org/10.1103/PhysRevD.78.106005}{\emph{Phys. Rev.
  D} {\bfseries 78} (Nov, 2008) 106005},
  [\href{https://arxiv.org/abs/0808.1725}{{\ttfamily 0808.1725}}].

\bibitem{Taylor1}
M.~Taylor, \emph{{Non-relativistic holography}},
  \href{https://arxiv.org/abs/0812.0530}{{\ttfamily 0812.0530}}.

\bibitem{Sachdev1}
J.~Crossno, J.~K. Shi, K.~Wang, X.~Liu, A.~Harzheim, A.~Lucas et~al.,
  \emph{Observation of the {D}irac fluid and the breakdown of the
  {Wiedemann-Franz} law in graphene},
  \href{https://doi.org/10.1126/science.aad0343}{\emph{Science} {\bfseries 351}
  (2016) 1058--1061}, [\href{https://arxiv.org/abs/1509.04713}{{\ttfamily
  1509.04713}}].

\bibitem{JacobsUndoped}
U.~G{\"u}rsoy, V.~Jacobs, E.~Plauschinn, H.~Stoof and S.~Vandoren,
  \emph{Holographic models for undoped {Weyl} semimetals},
  \href{https://doi.org/10.1007/JHEP04(2013)127}{\emph{Journal of High Energy
  Physics} {\bfseries 2013} (Apr, 2013) 127},
  [\href{https://arxiv.org/abs/1209.2593}{{\ttfamily 1209.2593}}].

\bibitem{Henningson1}
M.~Henningson and K.~Sfetsos, \emph{Spinors and the {AdS/CFT} correspondence},
  \href{https://doi.org/https://doi.org/10.1016/S0370-2693(98)00559-0}{\emph{Physics
  Letters B} {\bfseries 431} (1998) 63 -- 68},
  [\href{https://arxiv.org/abs/hep-th/9803251}{{\ttfamily hep-th/9803251}}].

\bibitem{Lee1}
S.-S. Lee, \emph{Non-{F}ermi liquid from a charged black hole: A critical
  {Fermi} ball}, \href{https://doi.org/10.1103/PhysRevD.79.086006}{\emph{Phys.
  Rev. D} {\bfseries 79} (Apr, 2009) 086006},
  [\href{https://arxiv.org/abs/0809.3402}{{\ttfamily 0809.3402}}].

\bibitem{Vegh1}
H.~Liu, J.~McGreevy and D.~Vegh, \emph{Non-{F}ermi liquids from holography},
  \href{https://doi.org/10.1103/PhysRevD.83.065029}{\emph{Phys. Rev. D}
  {\bfseries 83} (Mar, 2011) 065029},
  [\href{https://arxiv.org/abs/0903.2477}{{\ttfamily 0903.2477}}].

\bibitem{Vegh2}
T.~Faulkner, H.~Liu, J.~McGreevy and D.~Vegh, \emph{Emergent quantum
  criticality, {F}ermi surfaces, and {${\mathrm{AdS}}_{2}$}},
  \href{https://doi.org/10.1103/PhysRevD.83.125002}{\emph{Phys. Rev. D}
  {\bfseries 83} (Jun, 2011) 125002},
  [\href{https://arxiv.org/abs/0907.2604}{{\ttfamily 0907.2604}}].

\bibitem{Zaanen1}
M.~{\v C}ubrovi{\'c}, J.~Zaanen and K.~Schalm, \emph{String theory, quantum
  phase transitions, and the emergent {F}ermi liquid},
  \href{https://doi.org/10.1126/science.1174962}{\emph{Science} {\bfseries 325}
  (2009) 439--444}, [\href{https://arxiv.org/abs/0904.1993}{{\ttfamily
  0904.1993}}].

\bibitem{Landsteiner1}
K.~Landsteiner and Y.~Liu, \emph{The holographic {W}eyl semi-metal},
  \href{https://doi.org/https://doi.org/10.1016/j.physletb.2015.12.052}{\emph{Physics
  Letters B} {\bfseries 753} (2016) 453 -- 457},
  [\href{https://arxiv.org/abs/1505.04772}{{\ttfamily 1505.04772}}].

\bibitem{Landsteiner2}
K.~Landsteiner, Y.~Liu and Y.-W. Sun, \emph{Quantum phase transition between a
  topological and a trivial semimetal from holography},
  \href{https://doi.org/10.1103/PhysRevLett.116.081602}{\emph{Phys. Rev. Lett.}
  {\bfseries 116} (Feb, 2016) 081602},
  [\href{https://arxiv.org/abs/1511.05505}{{\ttfamily 1511.05505}}].

\bibitem{Jacobs1}
V.~P.~J. Jacobs, S.~J.~G. Vandoren and H.~T.~C. Stoof, \emph{Holographic
  interaction effects on transport in {Dirac} semimetals},
  \href{https://doi.org/10.1103/PhysRevB.90.045108}{\emph{Phys. Rev. B}
  {\bfseries 90} (Jul, 2014) 045108},
  [\href{https://arxiv.org/abs/1403.3608}{{\ttfamily 1403.3608}}].

\bibitem{Liu1}
Y.~Liu and Y.-W. Sun, \emph{{Topological nodal line semimetals in holography}},
   \href{https://arxiv.org/abs/1801.09357}{{\ttfamily 1801.09357}}.

\bibitem{Faulkner1}
T.~Faulkner and J.~Polchinski, \emph{Semi-holographic {Fermi} liquids},
  \href{https://doi.org/10.1007/JHEP06(2011)012}{\emph{Journal of High Energy
  Physics} {\bfseries 2011} (Jun, 2011) 12},
  [\href{https://arxiv.org/abs/1001.5049}{{\ttfamily 1001.5049}}].

\bibitem{Arpes}
U.~G{\"u}rsoy, E.~Plauschinn, H.~Stoof and S.~Vandoren, \emph{Holography and
  {ARPES} sum-rules},
  \href{https://doi.org/10.1007/JHEP05(2012)018}{\emph{Journal of High Energy
  Physics} {\bfseries 2012} (May, 2012) 18},
  [\href{https://arxiv.org/abs/1112.5074}{{\ttfamily 1112.5074}}].

\bibitem{HHH2}
S.~A. Hartnoll, C.~P. Herzog and G.~T. Horowitz, \emph{Holographic
  superconductors}, {\emph{Journal of High Energy Physics} {\bfseries 2008}
  (2008) 015}, [\href{https://arxiv.org/abs/0810.1563}{{\ttfamily 0810.1563}}].

\bibitem{Horowitz0}
G.~T. Horowitz and M.~M. Roberts, \emph{Zero temperature limit of holographic
  superconductors}, {\emph{Journal of High Energy Physics} {\bfseries 2009}
  (2009) 015}, [\href{https://arxiv.org/abs/0908.3677}{{\ttfamily 0908.3677}}].

\bibitem{Landsteiner3}
C.~Copetti, J.~Fern{\'a}ndez-Pend{\'a}s and K.~Landsteiner, \emph{Axial {H}all
  effect and universality of holographic {W}eyl semi-metals},
  \href{https://doi.org/10.1007/JHEP02(2017)138}{\emph{Journal of High Energy
  Physics} {\bfseries 2017} (Feb, 2017) 138},
  [\href{https://arxiv.org/abs/1611.08125}{{\ttfamily 1611.08125}}].

\bibitem{Horowitz1}
G.~T. Horowitz and M.~M. Roberts, \emph{Holographic superconductors with
  various condensates},
  \href{https://doi.org/10.1103/PhysRevD.78.126008}{\emph{Phys. Rev. D}
  {\bfseries 78} (Dec, 2008) 126008},
  [\href{https://arxiv.org/abs/0810.1077}{{\ttfamily 0810.1077}}].

\bibitem{Kovtun2}
P.~Kovtun and A.~Ritz, \emph{Universal conductivity and central charges},
  \href{https://doi.org/10.1103/PhysRevD.78.066009}{\emph{Phys. Rev. D}
  {\bfseries 78} (Sep, 2008) 066009},
  [\href{https://arxiv.org/abs/0806.0110}{{\ttfamily 0806.0110}}].

\bibitem{Plantz1}
N.~W.~M. Plantz, H.~T.~C. Stoof and S.~Vandoren, \emph{Order parameter
  fluctuations in the holographic superconductor}, {\emph{Journal of Physics B:
  Atomic, Molecular and Optical Physics} {\bfseries 50} (2017) 064001},
  [\href{https://arxiv.org/abs/1511.05112}{{\ttfamily 1511.05112}}].

\bibitem{Jacobs2}
V.~P.~J. Jacobs, S.~Grubinskas and H.~T.~C. Stoof, \emph{Towards a field-theory
  interpretation of bottom-up holography},
  \href{https://doi.org/10.1007/JHEP04(2015)033}{\emph{Journal of High Energy
  Physics} {\bfseries 2015} (Apr, 2015) 33},
  [\href{https://arxiv.org/abs/1411.4051}{{\ttfamily 1411.4051}}].

\bibitem{Jacobs3}
V.~P.~J. Jacobs, P.~Betzios, U.~G\"ursoy and H.~T.~C. Stoof,
  \emph{Electromagnetic response of interacting {Weyl} semimetals},
  \href{https://doi.org/10.1103/PhysRevB.93.195104}{\emph{Phys. Rev. B}
  {\bfseries 93} (May, 2016) 195104},
  [\href{https://arxiv.org/abs/1512.04883}{{\ttfamily 1512.04883}}].

\bibitem{Schnyder1}
A.~P. Schnyder, S.~Ryu, A.~Furusaki and A.~W.~W. Ludwig, \emph{Classification
  of topological insulators and superconductors in three spatial dimensions},
  \href{https://doi.org/10.1103/PhysRevB.78.195125}{\emph{Phys. Rev. B}
  {\bfseries 78} (Nov, 2008) 195125},
  [\href{https://arxiv.org/abs/0803.2786}{{\ttfamily 0803.2786}}].

\bibitem{Volovik1}
G.~E. Volovik, \emph{Topological invariants for standard model: From semi-metal
  to topological insulator},
  \href{https://doi.org/10.1134/S0021364010020013}{\emph{JETP Letters}
  {\bfseries 91} (Jan, 2010) 55--61},
  [\href{https://arxiv.org/abs/0912.0502}{{\ttfamily 0912.0502}}].

\bibitem{Volovik2}
M.~A. Silaev and G.~E. Volovik, \emph{Evolution of edge states in topological
  superfluids during the quantum phase transition},
  \href{https://doi.org/10.1134/S0021364012010110}{\emph{JETP Letters}
  {\bfseries 95} (Mar, 2012) 25--28},
  [\href{https://arxiv.org/abs/1108.1980}{{\ttfamily 1108.1980}}].

\bibitem{Policastro2}
B.~Dou\ifmmode~\mbox{\c{c}}\else \c{c}\fi{}ot, C.~Ecker, A.~Mukhopadhyay and
  G.~Policastro, \emph{Density response and collective modes of semiholographic
  non-{F}ermi liquids},
  \href{https://doi.org/10.1103/PhysRevD.96.106011}{\emph{Phys. Rev. D}
  {\bfseries 96} (Nov, 2017) 106011},
  [\href{https://arxiv.org/abs/1706.04975}{{\ttfamily 1706.04975}}].

\bibitem{Hartnoll1}
S.~A. Hartnoll, D.~M. Hofman and D.~Vegh, \emph{Stellar spectroscopy: Fermions
  and holographic {L}ifshitz criticality},
  \href{https://doi.org/10.1007/JHEP08(2011)096}{\emph{Journal of High Energy
  Physics} {\bfseries 2011} (Aug, 2011) 96},
  [\href{https://arxiv.org/abs/1105.3197}{{\ttfamily 1105.3197}}].

\bibitem{Iqbal1}
N.~Iqbal, H.~Liu and M.~Mezei, \emph{Semi-local quantum liquids},
  \href{https://doi.org/10.1007/JHEP04(2012)086}{\emph{Journal of High Energy
  Physics} {\bfseries 2012} (Apr, 2012) 86},
  [\href{https://arxiv.org/abs/1105.4621}{{\ttfamily 1105.4621}}].

\end{thebibliography}\endgroup

\end{document}